\documentclass[12pt,preprint]{aastex}
\usepackage{epsfig,graphicx}   
\def\ADNDT{{ At. Data Nucl. Data Tables}}

\def\PREP{{ Phys. Rep. }}

\def\rpro{{ r}-process}

\def\nn{$n_{n}$}
\def\t9{$T_{9}$}

\def\a{$\alpha$}    
        
\def\ye{$Y_{e}$}  

\def\yz{Y(Z,A)}
\def\yza{Y(Z,A+1)}

\def\beq{\begin{equation}}
\def\eeq{\end{equation}}
\def\beqar{\begin{eqnarray}}
\def\eeqar{\end{eqnarray}}

\def\gtaprx{ \mathrel{  \vcenter{
                        \offinterlineskip \hbox{$>$}
                        \kern 0.3ex \hbox{$\sim$}    } } }
\def\fun#1#2{\lower3.6pt\vbox{\baselineskip0pt\lineskip.9pt
  \ialign{$\mathsurround=0pt#1\hfil##\hfil$\crcr#2\crcr\sim\crcr}}}

\def\deg{{$^{\circ}$}}
\def\bd17{\mbox{BD +17\deg 3248}}
\def\cs22{\mbox{CS 22892-052}}

\def\deg{{$^{\circ}$}}

\shorttitle{r-Process in HEW of Core-Collapse SNe}
\shortauthors{Farouqi et al.}

\begin{document}

\title{Charged-Particle and Neutron-Capture Processes in the High-Entropy Wind of Core-Collapse Supernovae} 

\author{K. Farouqi\altaffilmark{1,2,3,4}, K.-L. Kratz\altaffilmark{4,5}, B. Pfeiffer\altaffilmark{4,5}, T. Rauscher\altaffilmark{6}, F.-K. Thielemann\altaffilmark{6} and J.W. Truran\altaffilmark{1,2,7}}

\altaffiltext{1}{Department of Astrophysics and Astronomy, University of Chicago, Chicago, IL 60637, USA; farouqi@uchicago.edu, truran@nova.uchicago.edu}

\altaffiltext{2}{Joint Institute for Nuclear Astrophysics (JINA), http://www.jinaweb.org}
\altaffiltext{3}{Institut f\"ur Kernchemie, Universit\"at Mainz, D-55128 Mainz, Germany}
\altaffiltext{4}{Virtual Institute for Nuclear Structure and Astrophysics (Vistars), D-55128 Mainz, Germany}
\altaffiltext{5}{Max-Planck-Institut f\"ur Chemie (Otto-Hahn-Institut), D-55128 Mainz, Germany; K-L.Kratz@mpic.de, BPfeiffe@uni-mainz.de}
\altaffiltext{6}{Department of Physics, University of Basel, 4056 Basel, Switzerland; Thomas.Rauscher@unibas.ch, F-K.Thielemann@unibas.ch}
\altaffiltext{7}{Physics Division, Argonne National Laboratory, Argonne, IL 60439, USA}





\begin{abstract}

The astrophysical site of the r-process is still
uncertain, and a full exploration of 
the systematics of this process in terms of its dependence on nuclear 
properties from stability to the neutron drip-line within realistic stellar
environments has still to be undertaken. Sufficiently high neutron to seed ratios can only be obtained either in 
very neutron-rich low-entropy environments or moderately neutron-rich 
high-entropy environments, related to neutron star mergers (or jets of 
neutron star matter) and the high-entropy wind of core-collapse supernova explosions.
As chemical evolution models seem to disfavor neutron star mergers, we focus
here on high-entropy environments characterized by entropy $S$, electron
abundance $Y_e$ and expansion velocity $V_{exp}$. We investigate the
termination point of charged-particle reactions, and we define a maximum
entropy $S_{final}$ for a given $V_{exp}$ and $Y_e$, beyond which the seed production of heavy elements fails due to the very small matter density. We then investigate whether an r-process subsequent to the charged-particle freeze-out can in principle be 
understood on the basis of the classical approach, which assumes a chemical 
equilibrium between neutron captures and photodisintegrations, possibly followed by a $\beta$-flow equilibrium. In particular, we illustrate how long such 
a chemical equilibrium approximation holds, how the freeze-out 
from such conditions affects the abundance pattern, and which role the late capture of neutrons originating from $\beta$-delayed neutron emission can play. Furthermore,  
we analyze the impact of nuclear properties from different theoretical mass models on the final abundances after these
late freeze-out phases and $\beta$-decays back to stability. 
As only a superposition of astrophysical conditions can provide a good
fit to the solar r-abundances, the question remains how such superpositions are attained, resulting in the apparently robust r-process pattern observed
in low metallicity stars.
\end{abstract}

\section{INTRODUCTION}
A rapid neutron-capture process (r-process) in an explosive scenario is
traditionally believed to be responsible for the nucleosynthesis of about half
of the heavy elements above Fe \citep{burb57,cam57}. While in recent years the
high-entropy wind (HEW) of core-collapse supernovae has been considered to be one of the most promising sites, hydrodynamical simulations still encounter difficulties to reproduce the astrophysical conditions under which this process occurs. Therefore, a model-independent approach, i.e., the so-called ``Waiting-Point'' approximation (WP), 
has been utilized for many years (see, e.g., \citep{ctt91,kratz93, kratz07}). This rather simple model, with the basic assumptions of an Fe-group seed, an $(n,\gamma)-(\gamma,n)$-equilibrium for constant neutron densities \nn\ at a chosen temperature $T$ over a process duration $\tau$ and an instantaneous freezeout, has helped to gain steadily improved insight into the systematics of an r-process in terms of its dependence on nuclear-physics input and astrophysical conditions.
Taking a specific seed nucleus (for convenience often $^{56}{\mbox{Fe}}$ 
was employed), the solar \rpro\ pattern with its three pronounced abundance
peaks at A=80, 130 and 195 can be
reproduced for a variation of neutron number densities \nn\ and a given 
temperature $T$. Whether the solar r-process residuals $\mbox{N}_{r,\odot}\simeq\mbox{N}_{\odot}-N_{s,\odot}$
is fully reproduced in each astrophysical event, i.e., whether each such
event encounters the full superposition of conditions required, is a matter of
debate (see, e.g., \citep{kratz93,wasserburg96,meyerbrown97,pfeiffer01a,snedencowan03,hondaaoki06,qian07,ott08,farouqi09a} and references therein).\\
In realistic astrophysical environments with time variations in \nn\ and
$T$, it has to be investigated whether at all and for which time
duration $\tau$ the supposed $(n,\gamma)-(\gamma,n)$-equilibrium of the
classical approach will hold and how freeze-out effects change this behavior. 
In general, late neutron captures may alter the final abundance distribution. 
In this case, neutron capture reactions will be important. Also $\beta$-delayed 
neutrons ($\beta$dn) can play a role in forming and displacing the 
abundance peaks after freeze-out.\\
An example of a more realistic astrophysical environment is the HEW expelled from newly formed (hot) neutron stars in
core-collapse supernovae, which continue to release neutrinos after the 
supernova shock
wave is launched~\citep{meyer93,qianwoosley96}. These neutrinos interact with
matter of the outermost proto-neutron star layers which are heated and ejected
in a continuous wind. The late neutrino flux also leads to moderately
neutron-rich matter~\citep{qianwoosley96} via interactions with neutrons and
protons and causes matter ejection with high entropies. However, from the
beginning problems were encountered to attain entropies sufficiently high in
order to produce the heaviest r-process nuclei (see, e.g., \citep{takahashi94,woosley94,thompson01,wanajo01,terasawa02}). Recent
hydrodynamic simulations for core-collapse supernovae support the idea that
these entropy constraints may be fulfilled in the late phase (after the initial explosion) when a reverse shock is forming~\citep{fryer06,arcones07,burrows07,janka07}.\\
The question is whether such high entropies occur at times with sufficiently high temperatures when an \rpro\ is still underway~\citep{wanajo07,kuroda08}.
Exploratory calculations to obtain the necessary conditions for an r-process
in expanding high-entropy matter have been undertaken by a number of groups
(see, e.g., \citep{meyer92,hoffman97,meyerbrown97,otsuki00,wanajo01,terasawa01,wanajo04,yoshida04,kuroda08}).
In the present calculations we focus on (a) the effects of different nuclear 
physics input (mass models FRDM (Finite Range Droplett Model,~\citep{moller95}, ETFSI-1 (Extended Thomas-Fermi with Strutinsky Integral),~\citep{aboussir95}, a version with quenching of shell closures proportional to the distance from stability ETFSI-Q,~\citep{pearson96}, the mass formula of Duflo \& Zuker (DUFLO-ZUKER),~\citep{duflo96} and a recent Hartree-Fock-Bogoliubov approach (HFB-17),~\citep{goriely09}), and (b) a
detailed understanding of the r-process matter flow far off stability, in particular testing
equilibria, freeze-out and effect of delayed neutron capture.
To investigate these effects we have applied a full network containing up to 
6500 nuclei and the corresponding nuclear masses, cross sections and 
$\beta$-decay properties. \\
Starting in the 1990's, in various publications discussing r-process nucleosynthesis, neutrino interactions were predicted to significantly affect the final r-process abundances. Neutrinos can affect r-process environments in four different ways at different stages:
(a) by determining the neutron-richness of matter ($Y_e$) via neutrino or 
antineutrino captures on neutrons and protons (see, e.g., \citep{meyer92}), (b) by destroying $\alpha$-particles by neutral current interactions leading to higher seed production, i.e., reducing the neutron to seed ratio (see, e.g., \citep{meyer95}), (c) by speeding up the flow to heavy nuclei by charged current reactions mimicing fast $\beta^-$-decays (see, e.g., \citep{fuller95,qianwoosley96} or finally (d) by acting via neutral-current (spallation) reactions to possibly ``filling up'' underabundances in the low-mass wings of the A$\simeq$130 and neutrino-induced fission (see, e.g., \citep{qian02,kolbe04}.
(b) acts like starting an r-process with a higher $Y_e$, (c) was found to be
rather unimportant~\citep{freiburghaus99}, in (d) the effects of 
$\beta$-delayed neutron emission seem to be dominant~\citep{kratz01,pfeiffer01a}, and
$\nu$-delayed fission was also found to be unimportant. Thus, our calculations are based on trajectories for the
matter density $\rho(t)$ and the temperature $T(t)$, originating from the HEW
expansions. An extended
parameter study of the \rpro\ has been performed in terms of entropy $S$,
electron abundance $Y_{e}$ and expansion velocity $V_{exp}$ of the hot bubble,
the latter quantity being related to the expansion timescale $\tau_{exp}$.
\section{THE COMPUTATIONAL FRAMEWORK: HIGH-ENTROPY EXPANSIONS AND  NUCLEAR PROPERTIES}
In the absence of self-consistent hydrodynamical models which also result in
r-process conditions permitting to produce matter all the way up to the third
peak and U and Th, we continue to utilize parametrized calculations in order
to survey the
dependence on nuclear properties and highlight a detailed understanding of 
the nucleosynthesis processing which has so far not been fully analyzed.
As mentioned in the introduction, many of such calculations have been performed
in the past which displayed final abundance distributions. We intend here 
to focus on monitoring
for which time intervals equilibria are obtained, for which nuclear
mass ranges this applies, and what are the effects of the emission and recapture of $\beta$-delayed neutrons on the (final) abundances.

We follow the description of adiabatically expanding homogeneous mass
zones as already utilized in~\citet{freiburghaus99}. 
Different mass zones have different
initial entropies so that the overall explosion represents a superposition
of entropies. The electron abundance $Y_{e}$, the entropy
$S$ and the expansion velocity $V_{exp}$ are parameters representing the degrees of freedom 
related to the expansion properties for the thermodynamic evolution 
and the nucleosynthesis in an adiabatic expansion.
\subsection{Thermodynamics}
\begin{figure}
\centerline{\psfig{file=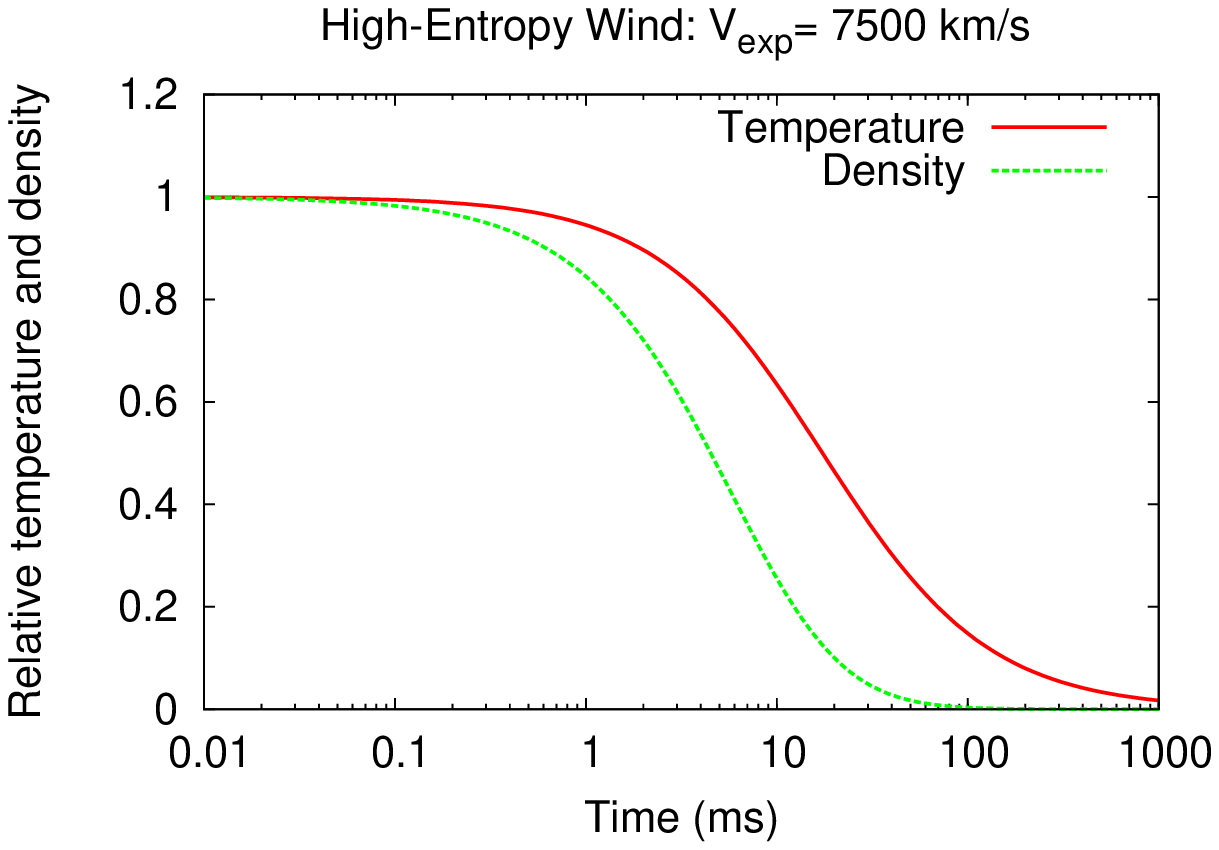,width=100mm,angle=00}}
\caption{Time evolution of temperature $T$ and density $\rho$ in the expanding
  hot bubble. The initial values of $T$ and $\rho$ are normalized to 1. The freeze-out of the charged-particle reactions occurs when the initial temperature in units of $10^{9}$ drops from $T_{9}=9$ to 3, and the freeze-out of a subsequent r-process occurs when $Y_{n}/Y_{r(A>4)}<1$.  
\label{fig1}}
\end{figure}

If the pressure per unit volume by relativistic particles (photons, electrons, 
positrons ..) is in excess of the total pressure due to non-relativistic 
particles, the entropy is dominated by radiation. This
occurs at high temperatures and moderate to small matter densities. As high 
temperatures also ensure a nuclear statistical equilibrium composition, it
is possible to choose an arbitary, but sufficiently high, initial temperature, e.g., $T_{9}\simeq 9$ ($T_9=T/10^9$ K). If the thermal energy $k_{B}T$ is larger than the energy
of the rest mass of electrons, we have also to
consider the contibutions by electrons and positrons. The radiation dominated
entropy is given by $S_{\gamma}=(4/3) a{T^{3}}/{\rho}$,
where $S_{\gamma}$ is the entropy per unit mass and $a$ is the
radiation constant. Making use of relations for ultrarelativistic fermions,
the electron and positron contributions are given by
$S_{e^{+},e^{-}}=({7}/{4})S_{\gamma}=({7}/{3}) a{T^{3}}/{\rho}$.
The total entropy per unit mass is therefore
\begin{equation}
S=S_{\gamma}+S_{e^{+},e^{-}}=
\frac{11}{3} a\frac{T^{3}}{\rho}, 
\end{equation} 
which is a factor $(11/4)$ larger than $S_{\gamma}$. After the
temperature has decreased to values where the corresponding energies become
comparable or smaller to the rest mass of electrons, the contributions of 
electrons and positrons can be neglected. The electrons become nonrelativistic 
and the positrons cease to exist. To consider both extreme
situations (pure radiation or radiation plus ultra-relativistic electrons and
positrons),~\citet{witti94} introduced a very close approximation for the entropy
\begin{equation}
S=S_{\gamma} \left[1+\frac{7}{4}f(T_{9})\right]
\end{equation} 
with
\begin{equation}
f(T_{9})=\frac{T^{2}_{9}}{T^{2}_{9}+5.3} .
\end{equation} 
The fit function $ f(T_{9})$ varies between 0 and 1. If we write the entropy
$S$ in units of k$_{B}$ per baryon and the density $\rho$ in
units of $10^{5}$ g cm$^{-3}$ ($\rho_{_{5}}$), the entropy $S$ can be expressed as
\begin{equation}\label{entropy}
S=1.21\frac{T^{3}_{9}}{\rho_{_{5}}}\left[1+\frac{7}{4}f(T_{9})\right].
\end{equation} 
Since the expansion of the hot bubble proceeds adiabatically
($TV^{1/3}=const$), the time evolution of the temperature is given by
\begin{equation}\label{temp}
T_{9}(t)=T_{9}(t=0)\left(\frac{R_{0}} {{R_{0}}+V_{exp}\, t} \right),
\end{equation}
where $R_{0}$ is the inital radius of an expanding sphere and  $V_{exp}$
is the expansion velocity of that sphere.
From Eq.~\ref{entropy} one can deduce the
matter density $\rho$ 
\begin{equation}\label{dichte}
\rho_{_{5}}(t)=1.21\frac{T^{3}_{9}}{S}\left[1+\frac{7}{4}f(T_{9})\right].
\end{equation}
Making use of a typical initial value $R_{0}=130$ km and a sufficiently high initial temperature $T_{9}=9$, which ensures nuclear statistical equilibrium to
consist only of neutrons and protons, we have performed calculations for a large 
grid of expansion velocities (1875, 3750, 7500, 15000 and 30000 km/s, for a given
$S$) and a large grid of entropies ($5\le S\le 415$). It should be noted that for the lowest entropies of this grid $(S<15)$ the 
above treatment is probably not valid, as the radiation dominance is 
not fulfilled. However, for the purpose of an r-process survey this 
parametrization seems still useful.\\
Fig.~\ref{fig1} shows the time evolution of $T_{9}(t)$ and 
$\rho_{_{5}}(t)$, both normalized to 1, for a selected expansion velocity of $V_{exp}=7500$ km/s. These normalized quantities are
identical for different entropies as long as $V_{exp}$ is kept constant. \\
The electron abundance $Y_{e}=\sum_{i} Z_{i}Y_{i}$ (with proton number
$Z_{i}$ and abundance $Y_{i}=X_i/A_i$ [mass fraction over mass number])
is equal to the averaged
electron or proton to nucleon ratio $<Z/A>$ and thus is a measure of the proton
to neutron ratio in the HEW. This is of key importance for the \rpro\ 
after the freeze-out of the charged-particle reactions. A neutron-rich wind 
means that $Y_{e}<0.5$. 
For an inititial $T_{9}=9$ only free protons and neutrons exist, leading to 
$1=X_{n}+X_{p}=Y_{n}+Y_{p}=Y_{n}+Y_{e}$. Therefore, a given $Y_e$ defines the
initial proton and neutron abundances $Y_p=Y_e$ and $Y_n=1-Y_e$.
The initial value of $Y_{e}$
depends ultimately on all weak interactions (including neutrino interactions)
with the available free nucleons in the HEW. While in the
early phase of core-collapse supernovae a $Y_{e}$ larger than 0.5 is expected
(see, e.g., \citep{frohlich06a,frohlich06b,pruet06}), the late phases are
expected to experience $Y_{e}<0.5$ and enable the onset of
an r-process. In order to test such conditions we have chosen a large grid of $Y_{e}$ (from 0.40 up to 0.499) for 
our calculations. \\ 

\subsection{Nuclear Networks and Nuclear Properties}

\subsubsection{The Charged-Particle Network}
The nucleosynthesis calculations during the early (hot phase in the) expansion 
of the wind until charged-particle freeze-out, were performed with 
the Basel nucleosynthesis code 
(see, e.g., \citep{thielemann96,brachwitz00,frohlich06a}; however, not including
neutrino interactions) making use of the
nuclear input as described there. Some new experimental rates were added
since, but for the majority of two-particle rates and their inverse reactions -- especially those for unstable nuclei -- the predictions of~\citet{rauscher00} (rate set FRDM) were used. The code includes all 
neutron and charged-particle induced reactions as well as their inverse reactions,
$\beta$-decays, electron captures, $\beta$dn emission and
particle transfer reactions. The charged-particle nuclear network considered for all simulations 
of this paper in the early expansion phase, before the onset of the r-process,
is shown in Table~\ref{table1}.
\subsubsection{The r-Process Network}
The \rpro\ calculations after charged-particle freeze-out made use
of the updated version of the code as documented in~\citet{freiburghaus99}, which includes 
neutron-induced reactions and $\beta$-decay properties as well as a simple
formulation for fission. The neutron capture and $\beta$-decay rates were
updated and modified in order to study the impact of a variety of mass models
(FRDM~\citep{moller95}, ETFSI-1~\citep{aboussir95}, ETFSI-Q~\citep{pearson96},
and HFB-17~\citep{goriely09}) and a mass formula
DUFLO-ZUKER~\citep{duflo96}. The size of the network was chosen specific to
each mass model because the neutron dripline is located at different mass
number $A$ for each mass model. Specific care was taken to include
$\beta$dn emission and their re-capture also during r-process freeze-out. Therefore, temperature-dependent neutron capture 
rates had to be introduced. For these purposes both the rate sets based on FRDM and ETFSI-Q of~\citet{rauscher00} were used for the experimentally unknown rates of neutron-rich nuclei. Details on the nuclear spectroscopy used in the calculation of these rates are given in~\citet{rauscher01}. To study the impact of further mass predictions by ETFSI-1, DUFLO-ZUKER and HFB-17, the effective neutron separation energies $S_n$ used in the computation of the photodisintegration rate, were modified according to the mass predicitions. \\
The method described here, making use of a splitting between the charged-particle
network (until charged-particle freeze-out) and a (fast) r-process network for
the neutron capture phase thereafter, is valid as long as the initial phase
(which includes charged particle reactions) is not producing nuclei beyond the
limit of the charged particle network (Z=46). This is the case for all
conditions in our parameter survey of $Y_e$, $S$, and expansion timescale. 
Calculations with very short expansion timescales~\citep{meyer02} differ, but we 
did not consider them here, as it has not been established that such conditions
reproduce the abundance pattern of heavy r-process nuclei, which is 
characterized by $\beta$-decay properties acting on longer timescales.
\section{EXPLOSIVE BURNING AND CHARGED-PARTICLE FREEZEOUT}
\begin{figure}
\centerline{\psfig{file=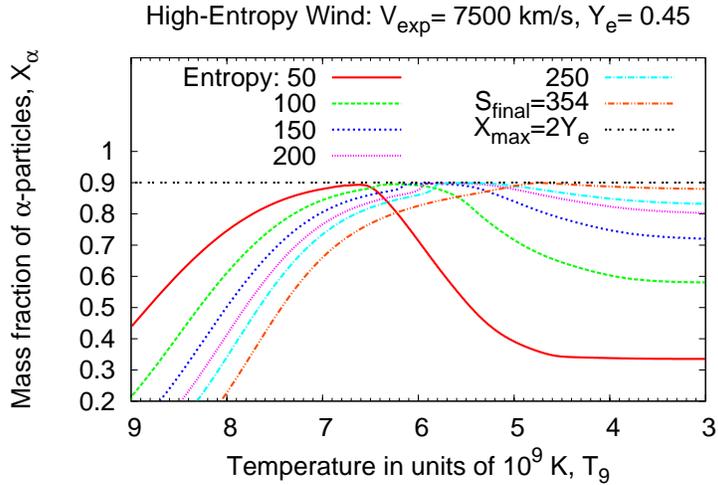,width=100mm,angle=00}}
\caption{Production and depletion of the $\alpha$-particles as a function of decreasing temperature. The maximum mass fraction of the $\alpha$-particles which can be reached is $X_{max}=2\, Y_{e}$. The depletion of the $\alpha$-particles depends both on the entropy $S$ and the electron abundance $Y_{e}$. Beyond a maximum entropy $S_{final}$, the production of any heavy seed is not possible because of the very small density. Hence, under such conditions an r-process is no longer possible.  
\label{fig2}}
\end{figure}

At initially high temperatures, \t9$\approx 9$, the baryonic matter
mainly consists of free neutrons and protons. Then
follows the recombination of the free nucleons into
\a-particles. The formation of \a-particles proceeds until the
temperature drops to values around \t9$\approx$ 6 when the
\a-particles dominate the nuclear abundances. The maximum abundance of the \a-particles which can be attained at that time, depends on \ye. If $Y_e<0.5$ and
all (initial) protons are consumed to form \a-particles, their abundance will be
\begin{equation}\label{ymaxalpha}
Y_{max}=\frac{Y_{p}}{2}=\frac{Y_{e}}{2},
\end{equation} 
and the corresponding mass fraction is 
\begin{equation}\label{xmaxalpha}
X_{max}=2 Y_{e}.
\end{equation} 
The mass fraction and the corresponding abundance of the remaining free neutrons is 
 \begin{equation}\label{xmaxneutrons}
X_{n}=Y_{n}=1-2 Y_{e}.
\end{equation} 
Table~\ref{table2} shows the maximum mass fraction $X_{max}$ and the corresponding $X_{n}$ before any seed can be synthesized, and Fig.~\ref{fig2} shows the behavior of the $\alpha$-particles for a selected $Y_{e}$-value and six different entropies.\\
 Between \t9$\approx 6$ and $3$, depending on the matter density, a complete or incomplete recombination of the \a-particles can take place. Thereby, the still existing neutrons will help  (a) to overcome the bottle-neck regions at A=5 and 8 and (b) to build neutron-rich seed nuclei.      
\subsection{The $\alpha$-Rich Freezeout}
\begin{figure}
\centerline{\psfig{file=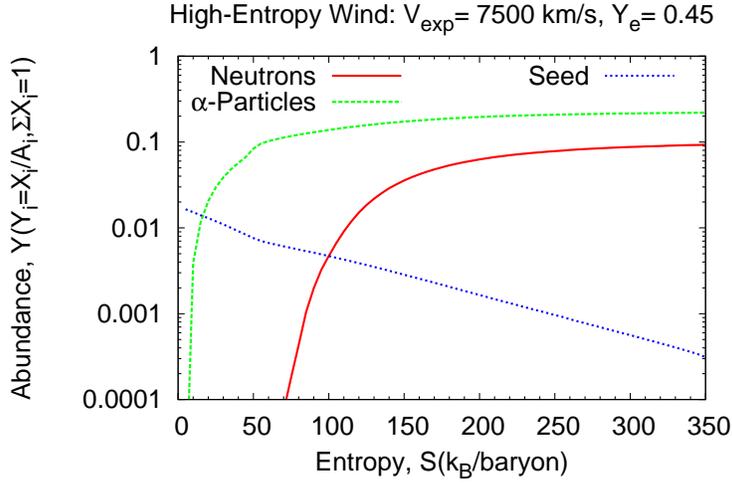,width=100mm,angle=0}}
\caption{Remaining abundances of $\alpha$-particles, heavy "seed" nuclei
and free neutrons after the charged-particle freeze-out as a function of the entropy for a selected electron abundance and expansion velocity. The r-process starts when the neutron to seed ratio becomes larger than unity, here at $S\approx 100$. 
\label{fig3}}
\end{figure}

When during the expansion of the hot bubble the temperature drops below
\t9$\approx 6$, the \a-particles start to
react with each other to form heavier nuclei via $^{12}{\mbox{C}}$, either
following the triple-\a\ -reaction
\begin{equation}
3\alpha\to {^{12}{\mbox{C}}},
\label{triple}
\end{equation} 
or, if neutrons are available, the reaction
\begin{equation}
\alpha+\alpha+n\to {^{9}{\mbox{Be}}},
\label{aan}
\end{equation} 
 followed by $^{9}\mbox{Be}(\alpha,n) ^{12}{\mbox{C}}$.\\
Both Eqs.~\ref{triple} \& \ref{aan} depend on the square of the 
matter density, and Eq.~\ref{aan} depends in addition on $Y_{e}$. They represent bottle-necks for the subsequent nucleosynthesis
towards heavier nuclei and are the reason for the well-known difference
between a normal and an $\alpha$-rich freeze-out of charged-particle
reactions. Further reactions of possible importance like $^{3}\mbox{He}+ ^{4}\mbox{He}\to^{7}\mbox{Be}+\gamma$, $^{7}\mbox{Be}+^{4}\mbox{He}\to ^{11}\mbox{C}+\gamma$, $^{4}\mbox{He}+^{2}\mbox{H}\to ^{6}\mbox{Li}+\gamma$, $^{6}\mbox{Li}+^{4}\mbox{He}\to ^{9}\mbox{Be}+^{1}\mbox{H}$, and $^{9}\mbox{Be}+^{4}\mbox{He}\to ^{12}\mbox{C}+\gamma$ are already included. In addition, test calculations with Fynbo's triple-alpha reaction rates~\citep{fynbo05} have been performed for two entropies $S=100$ and 250 k$_{B}$/baryon, although in several recent publications the data analysis with respect to the "missing" $2^{+}$ rotational-band member of the $^{12}$C Hoyle state and its astrophysical consequences have been questioned (see, e.g., \citep{freer09}). We also have tested
the NACRE triple-$\alpha$ rates for the above two entropies, and no significant differences in the seed contributions were observed at the charged-particle freeze-out. For high matter densities, i.e., low entropies ($S\propto T^3/\rho$), the
three-body reactions of Eqs.~\ref{triple} and~\ref{aan} will remain effective, leading
to a total recombination of \a-particles and free
  neutrons. The final abundances at charged-particle freeze-out close to
 \t9$\approx 3$ are dominated by iron group elements. For a moderate
neutron excess or $0.4<Y_e<0.5$, the free neutrons will be consumed completely
and no subsequent rapid neutron capture (r-process) will be possible. This is 
referred to as a normal freeze-out of charged-particle reactions (see, e.g., \citep{woosley94,freiburghaus99} and references therein).\\
In the case of lower matter densities, i.e., higher entropies, 
the reactions of Eqs.~\ref{triple} and~\ref{aan} cease to be effective for further
  recombination of \a-particles towards heavier nuclei. The degree to 
which such a production of heavier nuclei is prohibited is a gradual function
of $S$ and $Y_e$, ranging from \a-particle mass fractions of a few percent to
close to $100\, X_{max}=200\, Y_e$ \% (e.g., for $Y_e$=0.45 the maximum amount of the $\alpha$-particles is 90\%) and accordingly small amounts of heavy "seed" nuclei 
(see Fig.~\ref{fig3}). \\
In an \a-rich freeze-out the full nuclear statistical equilibrium
 (NSE) among all nuclei breaks down into quasi-equilibrium (QSE) subgroups
(see e.g., \citep{hix99}). For a strong \a-rich freeze-out and moderately
neutron-rich $Y_e$, the dominant abundances of heavy (seed) nuclei during the 
charged-particle freeze-out can be shifted to mass numbers 
between 80 and 110, thus overcoming the shell closure at N=50. 
The ratio of neutrons to heavy seed nuclei at this point is a function of
$Y_e$ (measuring the neutron excess) and entropy (determining the ratio of
heavy nuclei to \a-particles and indirectly also the amount of neutrons
consumed in these heavy nuclei). This can be seen from Fig.~\ref{fig3}
 and will be further elaborted in the following subsections. The neutron to seed ratio is a measure of the amount of neutron captures on 
seed nuclei and the resulting average mass number after an r-process, and 
indicates therefore the strength of the r-process. The fact that the seed abundances are dominated by nuclei
with mass numbers between 80 and 110, is {\it one} reason that an r-process
in the HEW can be faster than in classical calculations, which typically
starts below N=50.

\subsection{Freezeout Timescale}
We define the expansion time $\tau_{exp}$ between our arbitrarly chosen initial temperature $T_{9}= 9$ and its decrease to $T_{9}= 3$ as the timescale for the charged-particle reactions and as the time when the freeze-out of such reactions occurs within the expanding and cooling wind. According to 
Eq.~\ref{temp} one gets

\begin{equation}\label{tauexp}
\tau_{exp}=\frac{2 R_{0}}{V_{exp}},
\end{equation}
where $R_{0}$ is the initial radius of the hot bubble and $V_{exp}$ is its expansion velocity during the successful supernova explosion. For instance an expansion velocity of 7500 km/s corresponds to an expansion timescale of 35 ms. 
\subsection{Seed Distribution}
\begin{figure}
\centerline{\psfig{file=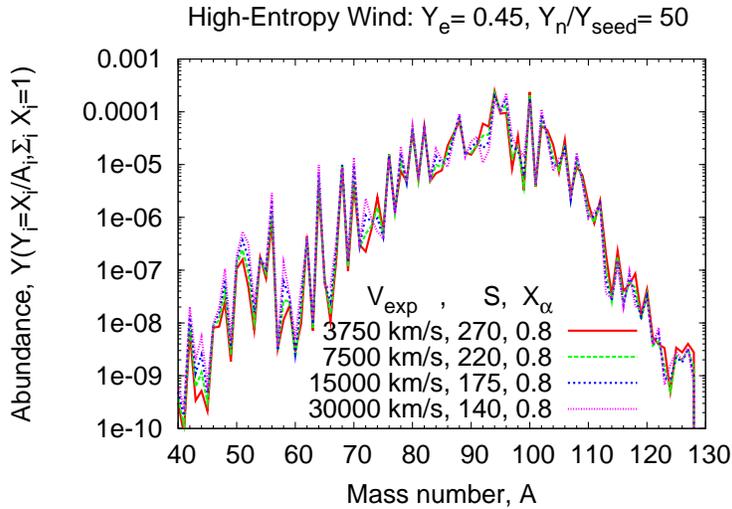,width=100mm,angle=00}}
\caption{The seed nuclei distributions for an arbitrary constant neutron to seed ratio of $Y_n/Y_{seed}=50$, an electron abundance $Y_{e}=0.45$ and for 4 different expansion velocities of the hot bubble, ranging from $V_{exp}=3750$ up to 30000 km/s. The corresponding entropies, ranging from $S=140$ up to 270 k$_B$/baryon are chosen such that the ratio mentioned above remains constant. The resulting seed distributions look very similar and are robust. The corresponding mass fractions of the $\alpha$-particles $X_{\alpha}$ are equal.      
\label{fig4}}
\end{figure}

\begin{figure}
\centerline{\psfig{file=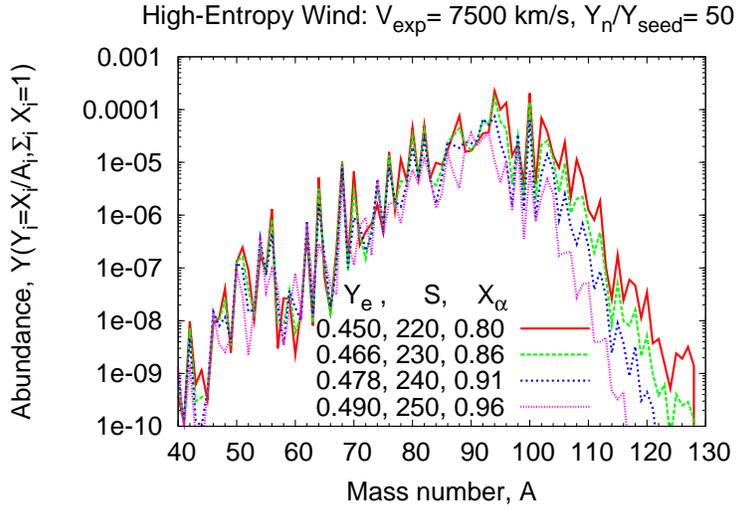,width=100mm,angle=00}}
\caption{The seed nuclei distributions for an arbitrary constant neutron to seed ratio of $Y_n/Y_{seed}=50$, an expansion speed $V_{exp}=7500$ km/s and 4 different electron abundances ranging from $Y_{e}=0.45$ up to 0.49. The corresponding entropies are chosen such that the ratio mentioned above remains constant. The resulting seed distributions are not robust. The corresponding mass fraction of the $\alpha$-particles decreases with decreasing electron abundance. 
\label{fig5}}
\end{figure}

After an $\alpha$-rich freeze-out, matter which surpassed the bottle-neck
 above A=4 is predominantly accumulated in nuclei of the Fe-group or
beyond. At sufficiently high entropies these nuclei are shifted to the mass 
numbers in the range $80\le\mbox{A}\le 110$. If free neutrons are still available, 
neutron captures on those seed nuclei can proceed after the freeze-out of the 
charged-particle reactions, thus leading to a subsequent r-process.
For a given $Y_{e}$, ($V_{exp},S$)-combinations
 can be chosen
which result in a fixed neutron to seed ratio $(Y_n/Y_{seed})$ and a unique 
distribution of seed nuclei at freeze-out (independent of the expansion 
timescale for such choices). This is shown in Fig.~\ref{fig4} for a selected $Y_{e}$-value with a constant neutron to seed ratio and for 4 different expansion 
velocities. This means that for an r-process with a given strength for a
specific $Y_e$, i.e., with a given initial neutron to seed ratio, both the
initial temperature $(T_{9}\approx 3)$ and the seed distribution are always
the same. However, in order to get the same r-process ejecta, i.e., in some
sense a universal r-rprocess, similar expansion velocities (timescales) are
required. The reason for this is that the r-process path for a given strength
also depends on the temperature profile which is related to the expansion
timescale. Similar to other authors \citep{arcones07,wanajo07}, we use the terms ``hot'' or ``cold'' 
r-process dependending on the average temperature experienced when the neutron
captures occur (which enables photodisintegrations to be effective or
not). For instance, a small expansion velocity ($V_{exp}<3000$ km/s) leads to
a ``hot'' r-process which is completely governed by an
$(n,\gamma)-(\gamma,n)$-equilibrium. On the other hand, higher expansion
velocities lead to a ``hybrid'' r-process, which is first equilibrium
dominated at high temperatures ($T_{9}\ge 1$), then followed by a
non-equilibrium phase at lower temperatures ($T_{9}<1$), i.e., the ``cold'' r-process where neutron captures and $\beta$-decays compete.
In Fig.~\ref{fig5} we show the dependence of the seed composition for the same wind characteristics (i.e., the same strength and timescale), but for different electron abundances $Y_{e}$. In contrast to the quite robust seed 
distribution of Fig.~\ref{fig4}, here we observe for a constant
$V_{exp}$ a significant variation with $Y_{e}$, with the major differences at the highest seed masses above A$\approx$ 90. Since the mass fraction of $\alpha$-particles increases with decreasing $Y_e$, the amount of the ejected heavy material $(X_{heavy}=1-X_{\alpha})$ is different for the same r-process strength, provided that the expelled matter contains roughly the same amount of mass
per entropy interval.

\subsection{Termination of the Charged-Particle Reactions}
There is a natural upper limit of the r-process-relevant entropies, i.e., the last entropy which still yields a significant amount of seed nuclei. We label those maximum entropies as $S_{final}$.
Beyond $S_{final}$ the recombination of the $\alpha$-particles, i.e., Eqs.~\ref{triple} \& \ref{aan} fail completely and the yields at such entropies only consist of $\alpha$-particles, free neutrons which decay to protons, and traces of $^{2}\mbox{H}$ ,$^{3}\mbox{H}$, $^{3}\mbox{He}$, $^{12}\mbox{C}$ and $^{16}\mbox{O}$. This is similar to the big bang nucleosynthesis where the dominant protons and the less abundant neutrons recombine to roughly 75$\%$ protons, 24$\%$ $^{4}\mbox{He}$ and traces of $^{6}\mbox{Li}$ and $^{7}\mbox{Be}$. Table~\ref{table3} shows an example of the yield of such a boundary entropy beyond $S_{final}$. In the limiting case of disappearing protons, the abundance and mass fraction of $\alpha$-particles and the neutrons are given by Eqs.~\ref{ymaxalpha} to~\ref{xmaxneutrons}, and are shown for some values of $Y_{e}$ in Table~\ref{table2}.\\
Table~\ref{table4} finally shows the results of simulations for two expansion velocities and a selection of $Y_{e}$-values. The second and the sixth column contain the initial entropies $(S_{initial})$ with a strength equal to unity, so that from the respective seed composition a subsequent r-process can occur. The third and seventh column contain the highest entropy $(S_{final})$, for which seed nuclei can be formed. Beyond $S_{final}$, no Fe-group seed nuclei are produced, and hence under such conditions an r-process is no longer possible. Obviously, $Y_e$-values larger than 0.49 will exhibit a depletion of the third peak elements and the actinides Th and U since the maximum $Y_n/Y_{seed}$ ratios are smaller than 100.\\
From Table~\ref{table4}, one can also see that for a given expansion speed, $S_{final}$ changes and increases as $Y_e$ decreases. This is because the bottle-neck reaction of Eq.~\ref{aan} additionally depends on $Y_e$, and - provided neutrons are available - it can proceed at densities smaller than those needed by the triple-\a\ reaction since an uncharged particle is involved.        
\section{THE R-PROCESS}
Starting with the conditions resulting from the charged-particle freeze-out 
presented in the previous sections,
we have performed our \rpro\ calculations with the code discussed in Section 2.
In addition to the original code version of~\citet{freiburghaus99}, we also have introduced temperature-dependent neutron capture rates. 
We have investigated the impact of different mass models and $\beta$-decay 
properties on the r-process duration and the final abundances after 
$\beta$-decay back to stability, and also have examined the role of 
$\beta$-delayed neutrons. Special emphasis was put on a detailed understanding when a chemical equilibrium between neutron captures and photo-disintegrations,
i.e., an $(n,\gamma)-(\gamma,n)$-equilibrium is fulfilled, when the freeze-out
from such conditions occurs and how neutron captures of the remaining free neutrons and
$\beta$-delayed neutrons can affect the final abundance pattern.

\subsection{General Results}
\begin{figure}
\centerline{\psfig{file=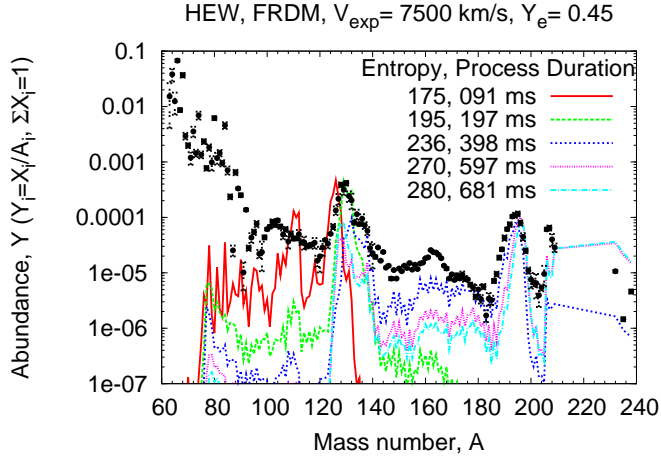,width=90mm,angle=00}}
\caption{A choice of five entropies which synthesize the mass region beyond A=110 by using the mass model FRDM~\citep{moller95} and the corresponding $\beta$-decay properties and neutron-capture rates. The solar r-abundances (black dots with error bars) are scaled such that the value of $^{130}$Te$_{r,\odot}$ coincides with the maximum of the $S=195$ curve at A=130. The process durations indicate the time between the freeze-out of the charged-particle reactions and the r-process. Note the depletion ot the REE mass region.
\label{fig8}}
\end{figure}

\begin{figure}
\centerline{\psfig{file=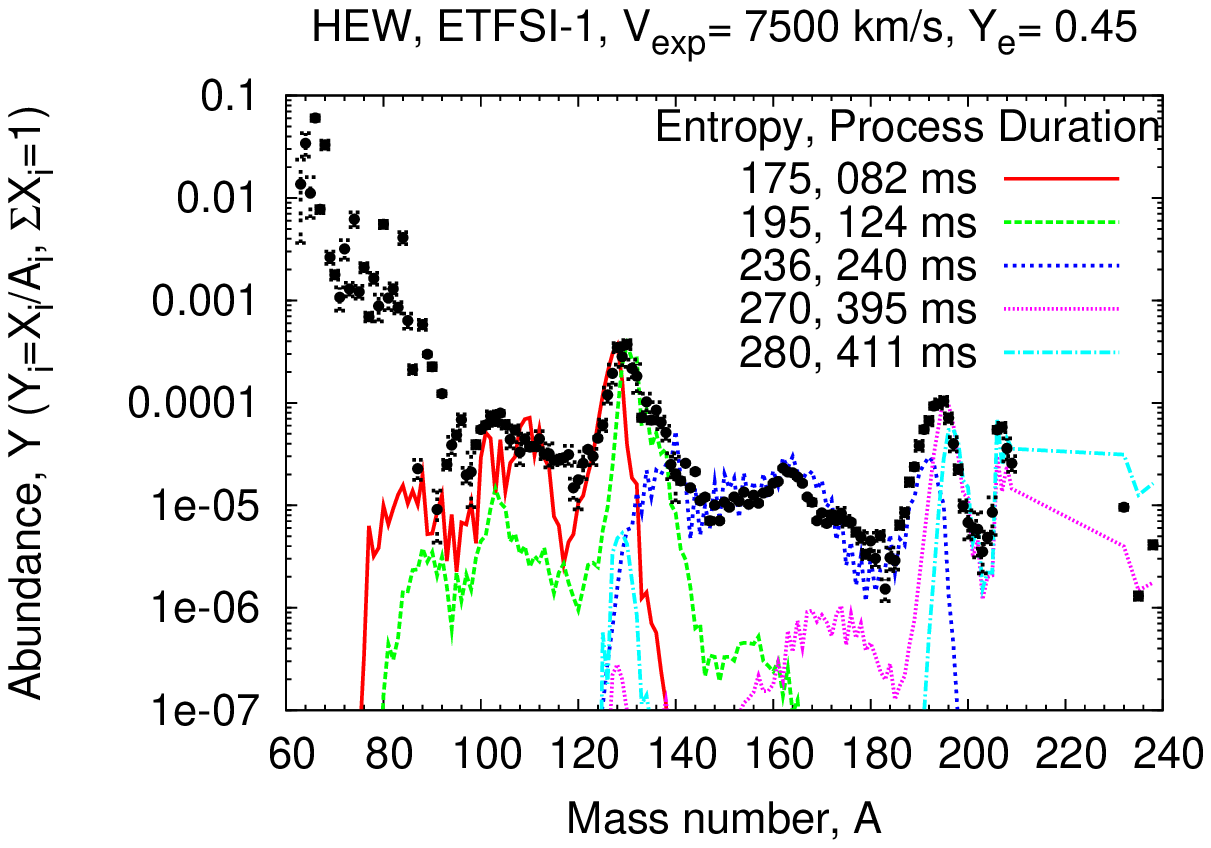,width=90mm,angle=00}}
\caption{A choice of five entropies which synthesize the mass region beyond A=110 by using the unquenched mass model ETFSI-1~\citep{aboussir95} and the corresponding $\beta$-decay properties and neutron-capture rates. See caption of Fig.~\ref{fig8} and text for further details.
\label{fig6}}
\end{figure}

\begin{figure}
\centerline{\psfig{file=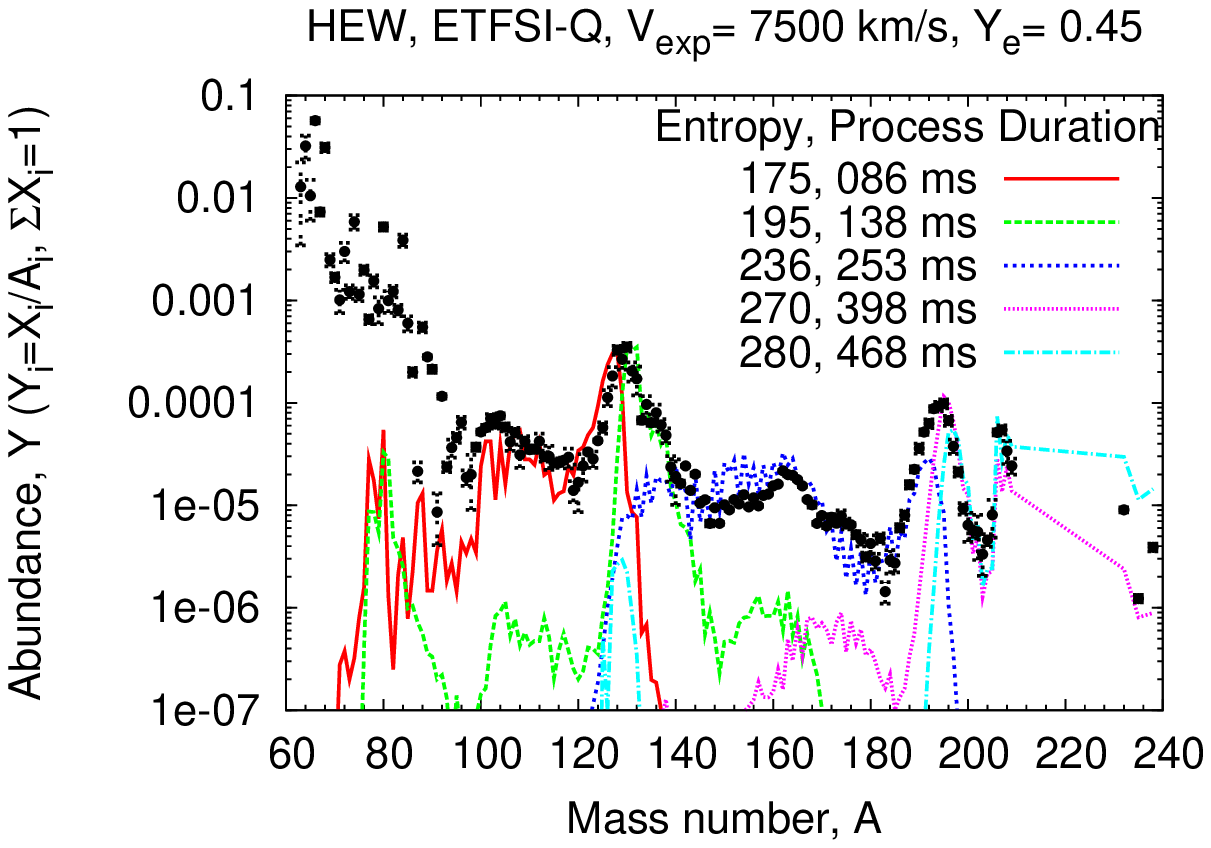,width=100mm,angle=00}}
\caption{A choice of five entropies which synthesize the mass region beyond A=110 by using the quenched mass model ETFSI-Q~\citep{pearson96} and the corresponding $\beta$-decay properties and neutron-capture rates. See caption of Fig.~\ref{fig8} and text for further details. 
\label{fig7}}
\end{figure}

\begin{figure}
\centerline{\psfig{file=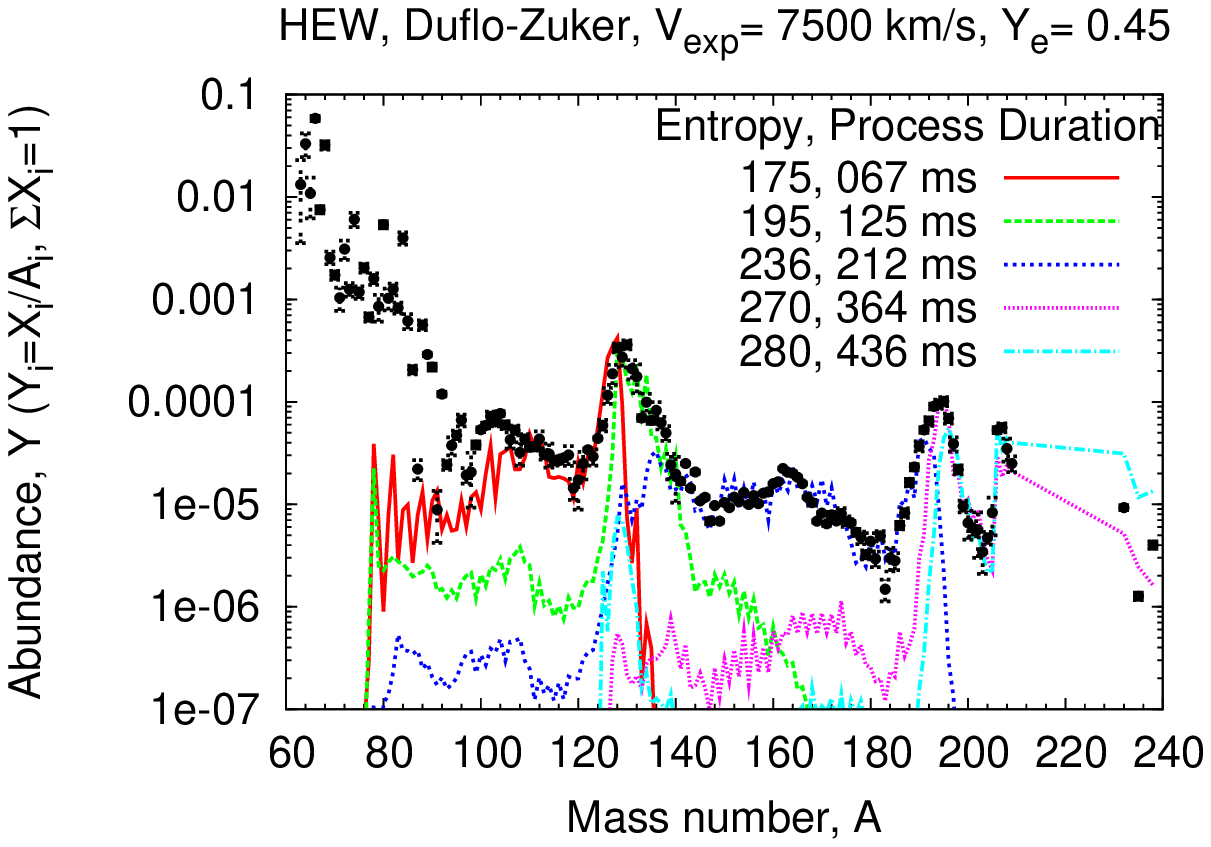,width=100mm,angle=00}}
\caption{A choice of five entropies which synthesize the mass region beyond A=110 by using the mass formula of DUFLO-ZUKER~\citep{duflo96} and the $\beta$-decay properties and neutron-capture rates calculated based on ETFSI-Q. See caption of Fig.~\ref{fig8} and text for further details. 
\label{fig9}}
\end{figure}

\begin{figure}
\centerline{\psfig{file=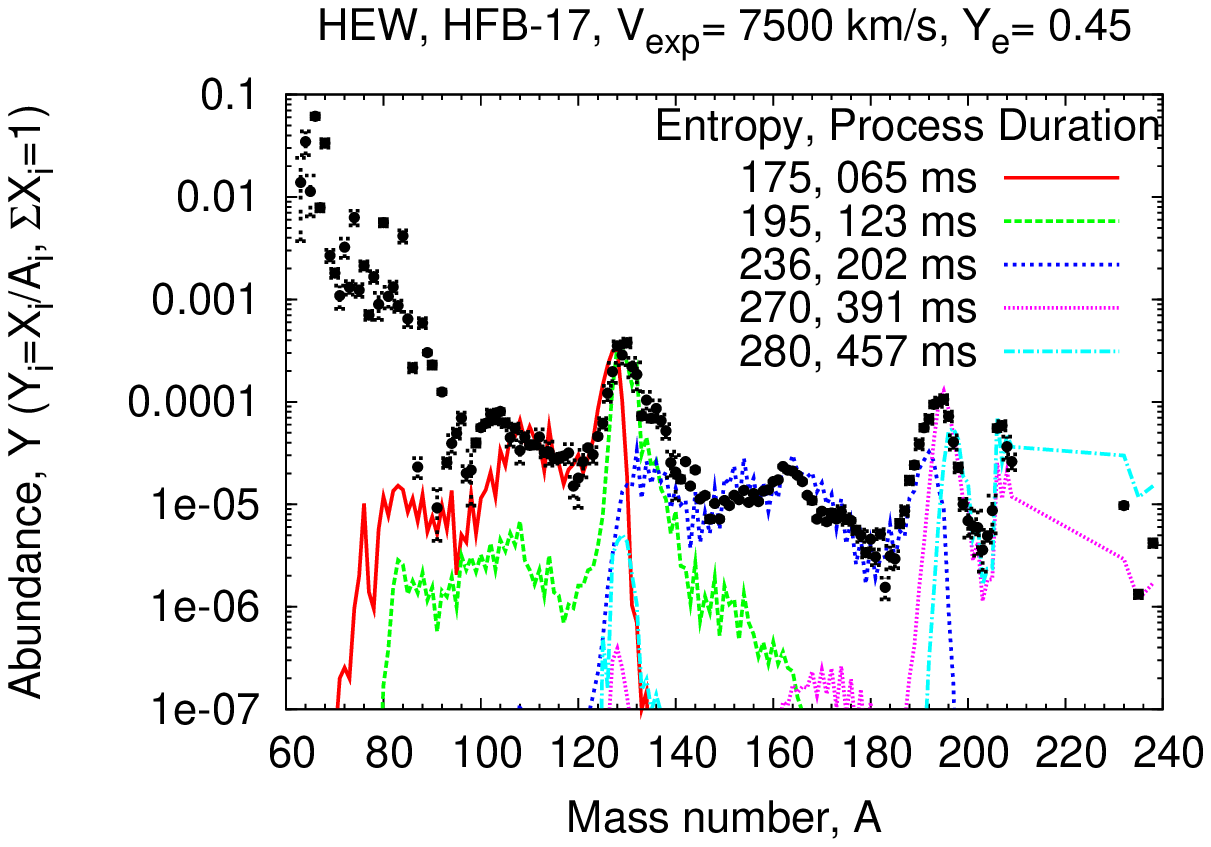,width=100mm,angle=00}}
\caption{A choice of five entropies which synthesize the mass region beyond A=110 by using the new mass model HFB-17~\citep{goriely09} and the $\beta$-decay properties and neutron-capture rates calculated based on ETFSI-Q. See caption of Fig.~\ref{fig8} and text for further details. 
\label{fig10}}
\end{figure}

Here we present a first survey of detailed network calculations for a variety 
of entropies, followed until the total consumption of neutrons. According to the entries in Table~\ref{table4}, entropies of
175, 195, 236, 270 and 280 k$_B$/baryon result in neutron to seed ratios
of about 23, 34, 66, 107 and 121. The first two components lead to
abundance patterns with maxima between A=100-130. The third component populates
the rare earth elements (hereafter, REE) in the mass range around A=162. The fourth component forms the abundances of the A=195 peak, and the fifth component proceeds to even heavier isotopes, also populating the actinides. These results are displayed in
Figs.~\ref{fig8} to \ref{fig10}, making use of five different mass 
models/formulae. Fission is here implemented in a simplified way, assuming complete and symmetric fission for nuclei with $A>260$. With a superposition of different entropy yields under the assumption that the ejected neutrino wind elements per equidistant entropy interval have equal volumes, final isotopic r-process abundances are obtained, which are displayed in Figs.~\ref{fig26} to \ref{fig28}. Deficiencies to the $N_{r,\odot}$ will be discussed in more detail later in section 5.2.

\subsection{Neutron Freeze-out}
While the r-process proceeds, the neutron abundance $Y_{n}$ decreases and the amount of the newly formed fresh r-process material increases as function of time. Since neutron captures destroy the seed nuclei, we replace the seed abundance $Y_{seed}$ by the r-process material abundance $Y_{r}$. With this, we define the neutron freeze-out as the instant when
 \begin{equation}\label{freeze1}
\frac{Y_{n}(t)}{Y_{r}(t)}<1,
\end{equation}    
and we label the corresponding time as $t_{freeze}$.
After $t_{freeze}$, the remaining neutrons and those produced by $\beta$delayed neutron emission will not change the final abundances dramatically. Figs.~\ref{fig8} to \ref{fig10} show for 4 mass models and one mass formula the freeze-out times $t_{freeze}$ for five selected entropies, i.e., $S=175, 195, 236, 270$ and  280, and the final abundance curves after $\beta$-decay back to stability. Table~\ref{table8} shows a comparison between two specifically selected mass models, i.e., FRDM and ETFSI-Q, for the same astrophysical conditions ($V_{exp}$, $Y_{e}$ and $S$). Obviously, the nuclear physics input seems to have a significant impact on the r-process abundances. For instance, with the FRDM mass model a longer process duration (by about a factor 1.5) is needed to produce the second $(A\simeq 130)$ and third $(A\simeq 195)$ r-process peaks than for ETFSI-Q. Similar differences in the time scales are also observed for ETFSI-1, DUFLO-ZUKER and HFB-17. These differences are mainly due to the strong $N=82$ shell closure of the FRDM mass model which in addition leads to an underproduction of the REE  mass region.\\
The weaker $N=126$ shell closure of FRDM causes an underproduction of Pb and an earlier onset of fission cycling than the other mass models. Hereafter, when speaking about freeze-out without any further specifications, the neutron freeze-out $(Y_n/Y_r<1)$ is meant.    
\subsection{Chemical Freeze-out}
\begin{figure}
\centerline{\psfig{file=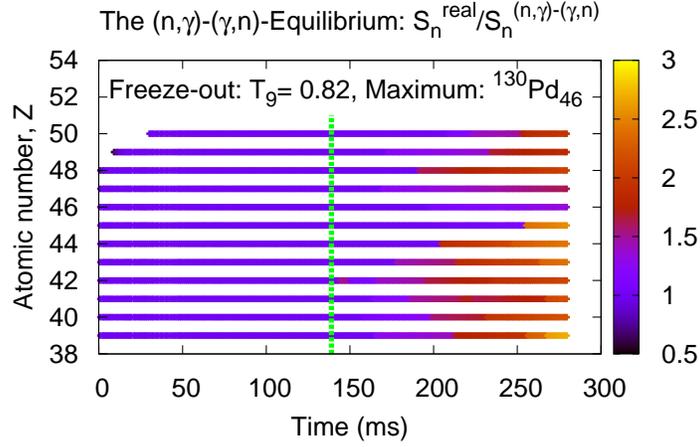,width=100mm,angle=00}}
\caption{The validity and endurance of the $(n,\gamma)-(\gamma,n)$-equilibrium for the A=130 peak formed by S=195 k$_{B}$/baryon. Color-coded is the ratio of the neutron separation energies $S_{n}^{real}(Z,A+1)$ of the right neighbor of the nucleus $
(Z,A)$ with the maximum abundance in the corresponding Z-chain, calulated by using the ETFSI-Q mass model and $S_{n}^{(n,\gamma)-(\gamma,n)}(Z,A+1)$ predicted by the nuclear Saha-equation in case of an existing chemical equilibrium between the nuclei $(Z,A)$ and $(Z,A+1)$. The equilibrium holds when $S_{n}^{real}/S_{n}^{(n,\gamma)-(\gamma,n)}\approx 1$. The vertical solid line shows the time of the r-process freeze-out. See text for further details.
\label{fig11}}
\end{figure}

\begin{figure}
\centerline{\psfig{file=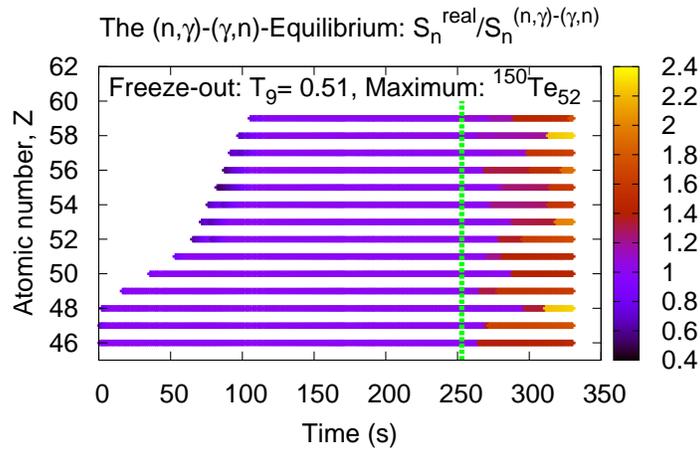,width=100mm,angle=00}}
\caption{The validity and the endurance of the $(n,\gamma)-(\gamma,n)$-equilibrium for the A=162 REE pygmy peak formed by $S=236$ k$_{B}$/baryon. See caption of Fig.~\ref{fig11} for further details. 
\label{fig12}}
\end{figure}

\begin{figure}
\centerline{\psfig{file=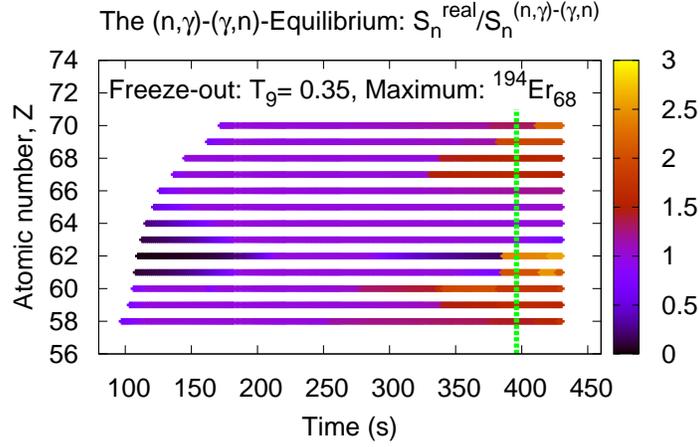,width=100mm,angle=00}}
\caption{The validity and the endurance of the $(n,\gamma)-(\gamma,n)$-equilibrium for the A=195 peak formed by $S=270$ k$_{B}$/baryon. See caption of Fig.~\ref{fig11} for further details. 
\label{fig13}}
\end{figure}

In case of a chemical equilibrium between neutron captures and reverse
photodisintegrations, or a so-called $(n,\gamma)-(\gamma,n)$-equilibrium
between neighboring nuclei $(Z,A)$ and $(Z,A+1)$ within an isotopic chain, the 
abundance ratio of these two nuclei is given by the nuclear Saha-equation
\begin{equation}\label{sahaeq}
\frac{\yza}{\yz}=\frac{n_{n}}{2}\frac{G(Z,A+1)}{G(Z,A)} \left(\frac{A+1}{A}\right)^{3/2}\left(\frac{2\pi\hbar^{2}}{m_{u}k_{B}T} \right)^{3/2}\mbox{exp}\left(\frac{-S_{n}(Z,A+1)}{k_{B}T}\right).
\end{equation} 
The  neutron separation energy $S_n$, the nuclear partition function
$G$ and the mass number $A$ represent the nuclear-physics input, whereas the
temperature $T$ and the neutron number density \nn\ represent the conditions of the
astrophysical environment. In the following, we want to analyze our r-process
calculations, testing if in high-entropy expansions an $(n,\gamma)-(\gamma,n)$-
equilibrium is achieved at all, how long it prevails and at which time
it breaks down. In addition, we will monitor the effect of abundance changes after
the freeze-out from such an equilibrium via final captures of remaining free neutrons and those from $\beta$-delayed netron emission.
If we express the product k$_{B}T$ in
units of \t9\ , Eq.~\ref{sahaeq} can be rewritten
as
\begin{equation}\label{sahaeq2}
\frac{\yza}{\yz}= 8.4\, 10^{-35}\,
n_{n}\, \frac{G(Z,A+1)}{G(Z,A)} \, \left(\frac{A+1}{A}\right)^{3/2}\,
T_{9}^{-3/2}\,\mbox{exp}\left(-\frac{11.604\, S_{n}(Z,A+1)}{T_{9}}\right).
\end{equation}
From Eq.~\ref{sahaeq2} we then obtain 
\begin{equation}\label{sneq}
S_{n}^{(n,\gamma-\gamma,n)}(Z,A+1)= -\frac{T_{9}}{11.604}\ln\left[1.2\, 10^{34}\, \frac{\yza}{\yz} \,\frac{G(Z,A)}{G(Z,A+1)}\, \left(\frac{A+1}{A}\right)^{-3/2}\,\frac{T_{9}^{3/2}}{n_{n}}\right].
\end{equation}
$S_{n}(Z,A+1)$ can also be calculated via
\begin{equation}\label{snmass}
S_{n}^{real}(Z,A+1)=m_{ex}^{Z,A+1}-(m_{ex}^{Z,A}+m_{ex}^{n}),
\end{equation}
where $m_{ex}^{Z,A+1}$, $m_{ex}^{Z,A}$ and $m_{ex}^{n}$ are the mass excesses of the nuclei
$(Z,A+1)$, $(Z,A)$ and the neutron, respectively.
We define the freeze-out with regard to the $(n,\gamma)-(\gamma,n)$-equilibrium as the instant, when the ratio of $S_{n}^{(n,\gamma-\gamma,n)}$ and $S_{n}^{real}$ of the nuclei $(Z,A+1)$ next to those lying in
the \rpro\ path $(Z,A)$ (which represent the abundance maxima in each isotopic chain
computed via relations given in Eqs.~\ref{sneq} and~\ref{snmass}
diverge from unity, and we label the corresponding time as $t_{chem}$.\\
Figs.~\ref{fig11} to~\ref{fig13} show the validity and duration of the $(n,\gamma)-(\gamma,n)$-equilibrium as a function of time for the entropies which synthesize the second, the REE pygmy and the third r-process abundance peaks. Obviously, the second peak is formed under a full chemical equilibrium. The ratios of the real neutron separation energies and those predicted by the Saha-equation are about unity for the whole Z-range (blue color). The break-out from the chemical equilibrium happens 60 ms later, i.e., clearly after the neutron freeze-out. Therefore, for the $A\simeq 130$ r-process peak one has $$t_{freeze}<t_{chem}\, .$$ The broad REE pygmy peak with its abundance maximum around $A\simeq 162$ is formed at the limit of a chemical equilibrium, where $$t_{freeze}\approx t_{chem}\, .$$ However, the $A\simeq 195$ peak is no longer built under chemical equilibrium conditions. This means $$t_{freeze}>t_{chem}\, .$$ Therefore, the morphology of the rapid neutron-capture process in the neutron-rich HEW can be depicted as a combination of three types:
\begin{enumerate}
\item a ``hot'' r-process, which proceeds and ends already at relatively high temperatures $(T_9\approx 0.8)$, maintaining a full chemical equilibrium until the $A\simeq 130$ peak is formed.
\item a ``hybrid'' r-process, which proceeds at high temperatures and ends at low temperatures $(T_9\approx 0.5)$, maintaining a partial chemical equilibrium until the REE pygmy peak is formed.
\item a ``cold'' r-process, which proceeds at high temperatures and ends at
  even lower temperatures $(T_9\approx 0.35)$ with no chemical equilibrium until
  the $A\simeq 195$ peak is formed (see also~\citep{arcones07,wanajo07}).
\end{enumerate}   
Thus, for the latter two r-process types the neutron capture rates will play a significant role for the mass region beyond $A\simeq 140$.
\subsection{Dynamical Freeze-out}
The dynamical freeze-out occurs when the neutron depletion becomes inefficient, and thus the timescale $\tau_{n}$ is larger than the hydrodynamical timescale $\tau_{hyd}$, i.e., 
\begin{equation}\label{tdyn}
\left|\frac{Y_n}{\dot{Y}_n}\right|> \left|\frac{\rho(t)}{\dot{\rho}(t)}\right|.
\end{equation}
We label the corresponding time with $t_{dyn}$. Then, $\Delta t =t_{dyn}-t_{freeze}$ is the time interval when the last free neutrons are captured and $\beta$-delayed neutrons are recaptured during the decay back to stabillity. Depending on the capability of the radioactive progenitor nuclei to emit $\beta$-delayed neutrons after $t_{freeze}$, the freeze-out times $t_{dyn}$ and $t_{freeze}$ may be identical or can differ from each other. Beyond $t_{dyn}$ the neutron number density $n_n$ drops to values $\le 10^{18} \,\mbox{n/cm}^3$, thus representing the signature for the end of the r-process.
\subsection{The $\beta$-Flow Equilibrium} 
\begin{figure}
\centerline{\psfig{file=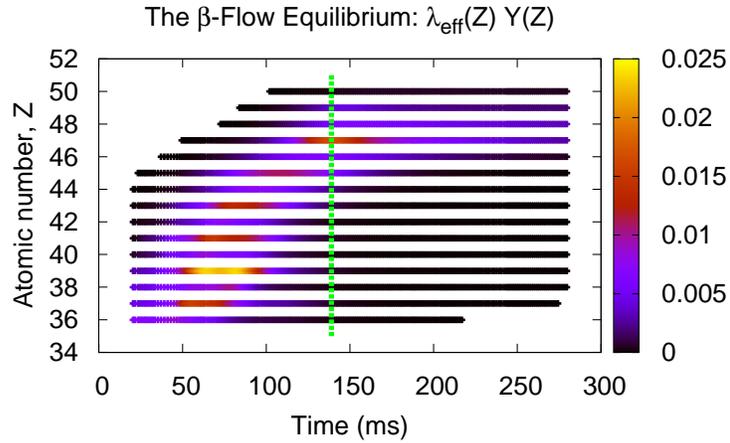,width=100mm,angle=00}}
\caption{The validity and endurance of the $\beta$-flow equilibrium for the A=130 peak formed by $S=195$ k$_{B}$/baryon as a function of time. Color-coded is the quantity $\lambda_{\mbox{eff}}(Z) Y(Z)$ of each Z-chain. The vertical line shows the time of the r-process freeze-out. Here, a global $\beta$-flow equilibrium does not occur, but local equilibria do appear between the populated odd-Z and even-Z chains before the r-process freezes out. After the freeze-out the $\beta$-flow equilibrium proceeds steadily to higher Z-chains (Z=47 and 48) around $t\simeq 150$ ms, and proceeds further to Z=50 before it completely breaks out.  
\label{fig14}}
\end{figure}

\begin{figure}
\centerline{\psfig{file=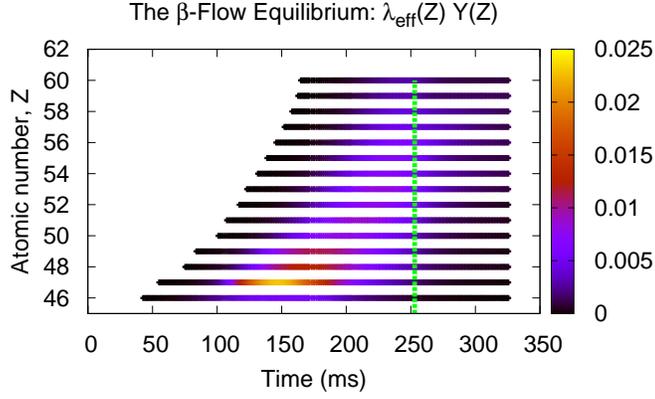,width=90mm,angle=00}}
\caption{The validity and endurance of the $\beta$-flow equilibrium for the REE pygmy peak formed by $S=236$ k$_{B}$/baryon as a function of time. Prior to the freeze-out, a wide-range equilibrium occurs between the Sn (Z=50) and Nd (Z=60) Z-chains when the matter flow passes the In (Z=49) Z-chain at $t\approx 200$ ms. After the freeze-out, the equilibrium lasts about 50 ms between the most populated Z-chains in the REE region. See caption of Fig.~\ref{fig14} and text for more details.  
\label{fig15}}
\end{figure}

\begin{figure}
\centerline{\psfig{file=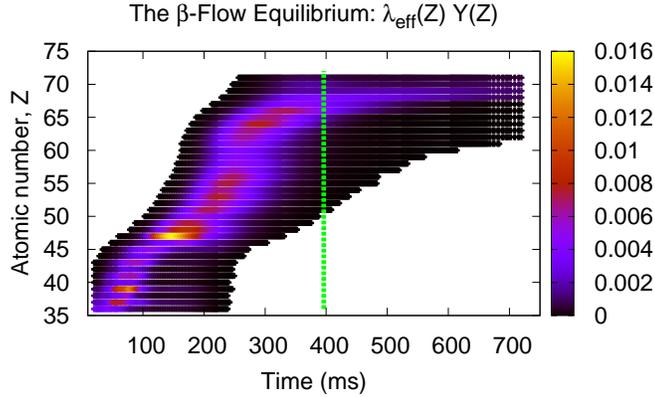,width=90mm,angle=00}}
\caption{The validity and endurance of The $\beta$-flow equilibrium for the A=195 peak formed by $S=270$ k$_{B}$/baryon as a function of time. Prior to the freeze-out, a strong local equilibrium occurs between the Sm (Z=62) and Er (Z=68) Z-chains when the mater flow passes the Pm (Z=61) Z-chain at $t\approx 250$ ms. After the freeze-out, a weak equilibrium lasts for about 300 ms between the Er (Z=68), Tm (Z=69) and Yb (Z=70) Z-chains. See caption of Fig.~\ref{fig14} and text for more details.
\label{fig16}}
\end{figure}

 During the r-process the abundance flow from each isotopic chain to the next is governed by $\beta$-decays. We can define a total abundance in each isotopic chain $Y(Z)=\sum_{A}Y(Z,A)$, and each $Y(Z,A)$ can be expressed as $Y(Z,A)=Y(Z)P(Z,A)$, where $P(Z,A)$ represents the population coefficient of the nucleus (Z,A). In case of a steady flow of $\beta$-decays between
isotopic chains, in addition to an $(n,\gamma)-(\gamma,n)$-equilibrium, we also have
\begin{equation}\label{steady}
\sum_{A}Y(Z,A)\lambda_{\beta}(Z,A)=Y(Z)\sum_{A}P(Z,A)\lambda_{\beta}(Z,A)    =Y(Z)\lambda^{eff}_{\beta}(Z)=const.
\end{equation}
In other words, each Z-chain can be treated as an effective nucleus with a $\beta$-decay rate $\lambda^{eff}$. Under such conditions the assumption of an abundance $Y(Z_{min})$ at a minimum Z-value would be sufficent to predict the whole set of abundances as a function of A, provided that an $(n,\gamma)-(\gamma,n)$-equilibrium and a steady flow has been reached. Already in the past, various groups have discussed the possibility of either a global r-process steady-flow equilibrium or local $(n,\gamma)-(\gamma,n)$ and $\beta$-flow equilibria in between neutron shell closures (see, e.g., \citep{burb57,seeger65,hillebrandt76,cam83a,cam83b}), assuming different astrophysical sites and different theoretical nuclear-physics input. First experimental evidence for a combination of an $(n,\gamma)-(\gamma,n)$-equilibrium and a local steady flow at N=50 and N=82 came from the work of \citet{kratz86,kratz88,kratz93} in their site-independent approach. This result that the $\beta$-flow is not global was shortly afterwards confirmed by \citet{meyer93} for the specific site of a high-entropy hot bubble. In the present paper, we extend the above steady-flow studies and show under which entropy and temperature conditions which type of equilibrium is reached and when it breaks down. \\
In Figs.~\ref{fig14}-\ref{fig16} we show the quantity $Y(Z)\lambda^{eff}_{\beta}(Z)$ (color-coded) as a function of time for entropies of $S=195$, 236 and 270 which produce the second peak, the REE pygmy peak and the third r-process peak regions, respectively. Those entropies were selected because they indicate best how the $\beta$-flow proceeds through both shell closures (N=82 and 126) and the REE region between them.  As a function of time, the following conclusions can be drawn: 
\begin{itemize}
\item Before approaching the freeze-out, a $\beta$-flow equilibrium is established for
  isotopic chains with Z$<$43, i.e., where the abundance distribution does not
  populate the N=82 shell closure. 
\item After reaching the freeze-out, another late $\beta$-flow proceeds through the most populated Z-chains: (a) Z=46-49 for the A$\simeq$130 mass region, (b) Z=50-63 for the REE pygmy peak region and (c) Z=64-70 for the A$\simeq$195 mass region.
\end{itemize}
Thus, we confirm the earlier conclusions (see, e.g., \citep{kratz93,meyer93,thielemann94,pfeiffer01a}), that in the r-process a global $\beta$-flow equilibrium does not occur, but local equilibria are established in between neutron shell closures where short $\beta$-decay half-lives are encountred. These equilibria are essential for a succesful reproduction of the N$_{r,\odot}$ distribution (see Chapter 5). 
\subsection{The Progenitors of $^{130}$Te, $^{162}$Dy and $^{195}$Pt, and the Role of $\beta$-Delayed Neutrons}
\begin{figure}
\centerline{\psfig{file=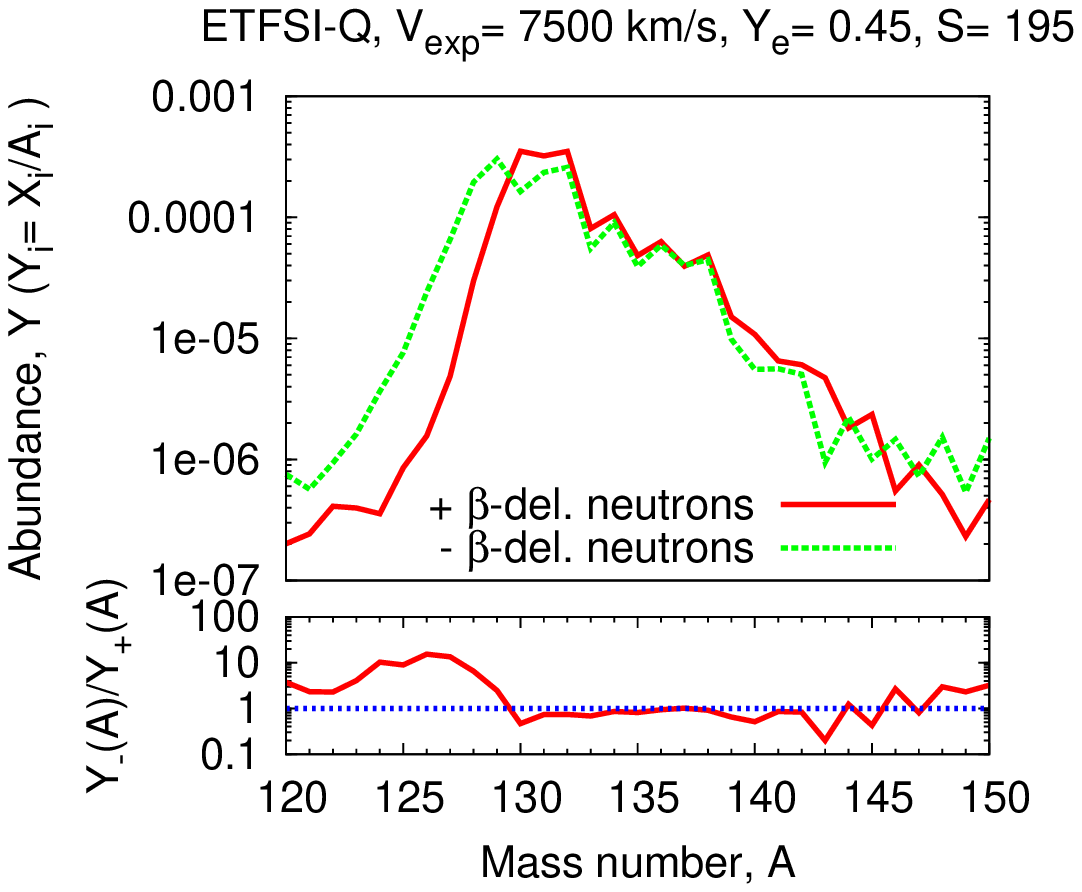,width=100mm,angle=00}}
\caption{The upper panel shows two abundance distributions for the same astropysical parameters $(S,Y_e,V_{exp})$ with $(+\beta)$ and without $(-\beta)$ recapturing of the previously emitted $\beta$dns. Their recapture causes the maximum abundance to shift from A=129 to A=130. The lower panel shows the ratio of the isotopic abundances for both cases, without and with $\beta$dns. See text for further details. 
\label{fig17}}
\end{figure}

\begin{figure}
\centerline{\psfig{file=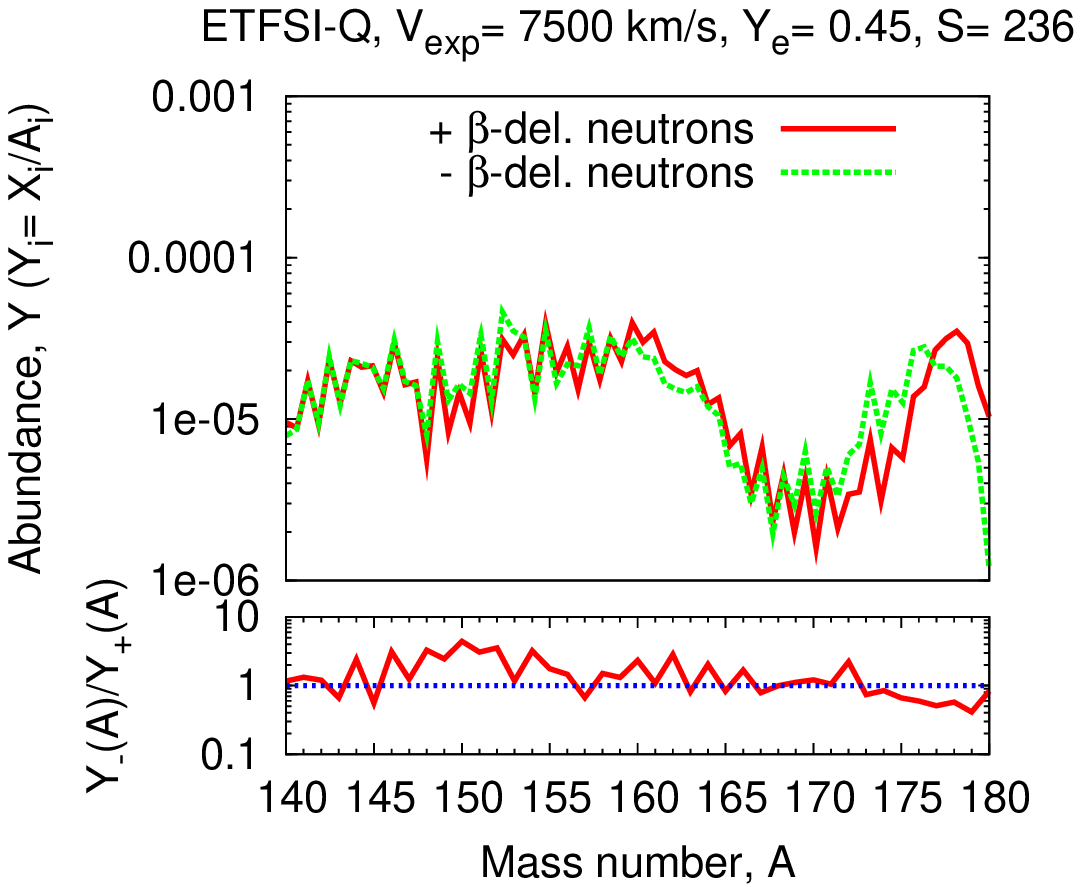,width=100mm,angle=00}}
\caption{The upper panel shows two abundance distributions for the same astropysical parameters $(S,Y_e,V_{exp})$ with $(+\beta)$ and without $(-\beta)$ recapturing of the previously emitted $\beta$dns. Their recapture causes the gradual shift of the maximum abundance to A$\simeq$162. The lower panel shows the ratio of the isotopic abundances for both cases, without and with $\beta$dns. See text for further details.  
\label{fig18}}
\end{figure}

\begin{figure}
\centerline{\psfig{file=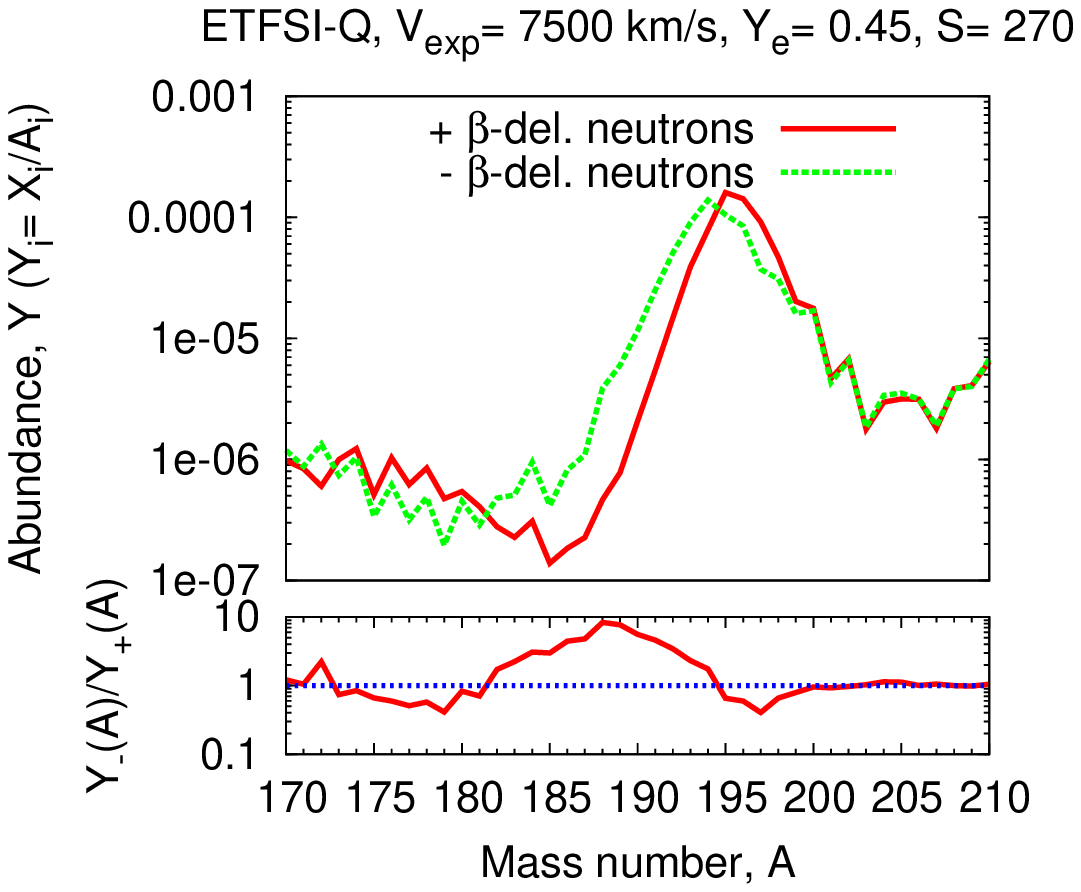,width=100mm,angle=00}}
\caption{The upper panel shows two abundance distributions for the same astropysical parameters $(S,Y_e,V_{exp})$ with $(+\beta)$ and without $(-\beta)$ recapturing of the previously emitted $\beta$dns. Their recapture causes the gradual shift of the maximum abundance from A=194 to A=195. The lower panel shows the ratio of the isotopic abundances for both cases, without and with $\beta$dns. See text for further details.    
\label{fig19}}
\end{figure}

\begin{figure}
\centerline{\psfig{file=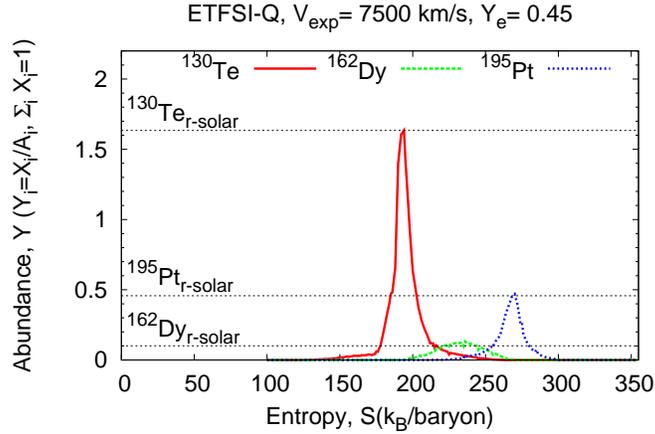,width=90mm,angle=00}}
\caption{The calculated abundances of $^{130}$Te, $^{162}$Dy and $^{195}$Pt as a function of the entropy $S$ by using the quenched mass model ETFSI-Q. The three curves are scaled with the same factor to normalize the maximum of the $^{130}$Te curve to the abundance of $^{130}$Te$_{r,\odot}$. Note the good agreement of the the relative heights of the peak maxima with the corresponding solar r-abundances in case of a continuous ejection of neutrino wind elements with equal volumes per equidistant entropy interval $(V(S)=const)$.
\label{fig20}}
\end{figure}

\begin{figure}
\centerline{\psfig{file=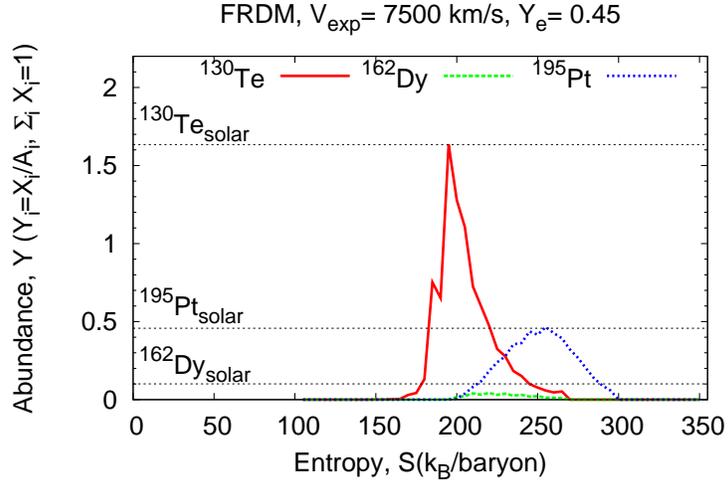,width=100mm,angle=00}}
\caption{See Fig.~\ref{fig20}. The mass model used here is FRDM. Again, note the good agreement of the maximum of the $^{195}$Pt curve with its corresponding solar r-abundance. The underproduction of the REE mass region is due to the strong N=82-shell closure of the mass model used. 
\label{fig21}}
\end{figure}

\begin{figure}
\centerline{\psfig{file=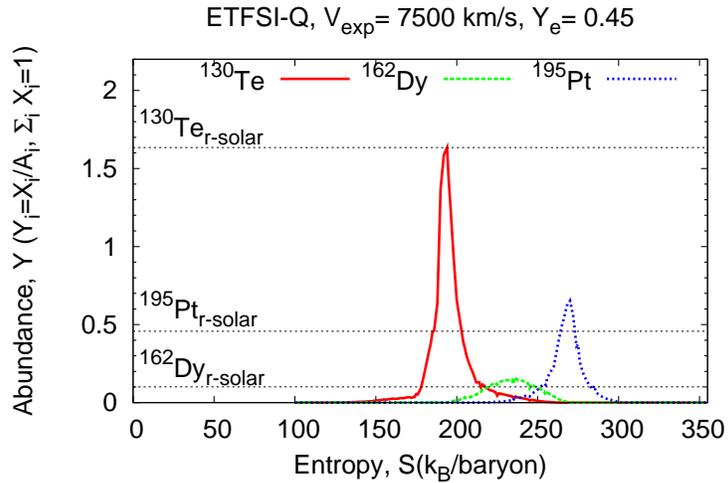,width=100mm,angle=00}}
\caption{See Fig.~\ref{fig20}, but with the ejection of neutrino-wind elements with equal masses per equidistant entropy interval $(M(S)=const)$. Note the slight disagreement of the top of the $^{162}$Dy and $^{195}$Pt curves with their corresponding solar r-abundances. 
\label{fig22}}
\end{figure}

\begin{figure}
\centerline{\psfig{file=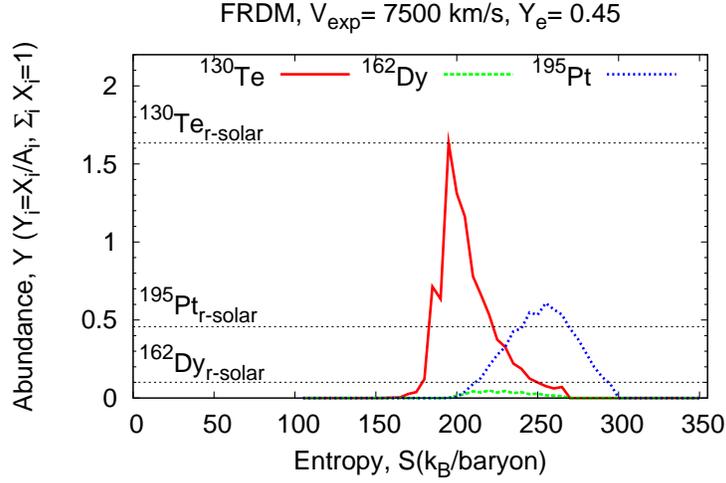,width=100mm,angle=00}}
\caption{See Fig.~\ref{fig20}, but with the ejection of neutrino-wind elements with equal masses per equidistant entropy interval $(M(S)=const)$ and by using the FRDM masses. Note the strong underproduction of $^{162}$Dy and the slight disagreement of the top of the $^{195}$Pt curve with its corresponding solar r-abundance.   
\label{fig23}}
\end{figure}

\begin{figure}
\centerline{\psfig{file=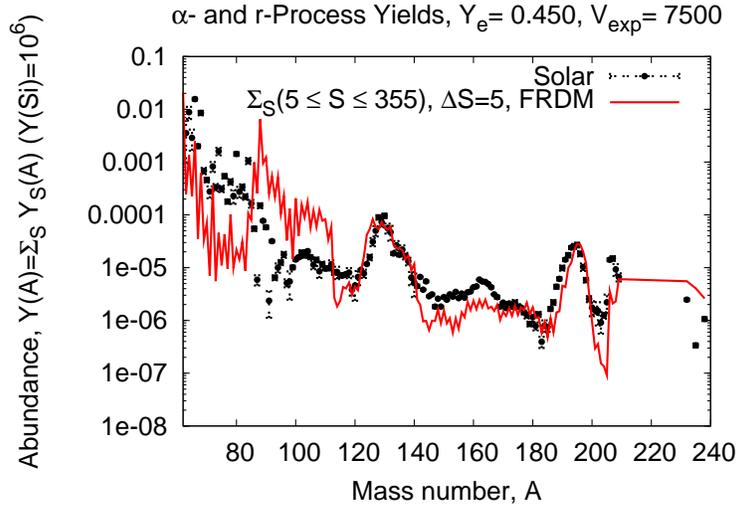,width=100mm,angle=00}}
\caption{Superposition of entropies from $S=5$ up to the maximum entropy $S_{final}=355$ using the mass model FRDM. The overall fit of the mass region from A=110 to 238 is good, except for the REE mass region which is strongly underproduced by a factor of 3. Furthermore, the left wing of the A=195 peak is displaced and an abundance trough occurs at A$\approx$205 which leads to a strong underproduction of the Pb mass region.
\label{fig26}}
\end{figure}

\begin{figure}
\centerline{\psfig{file=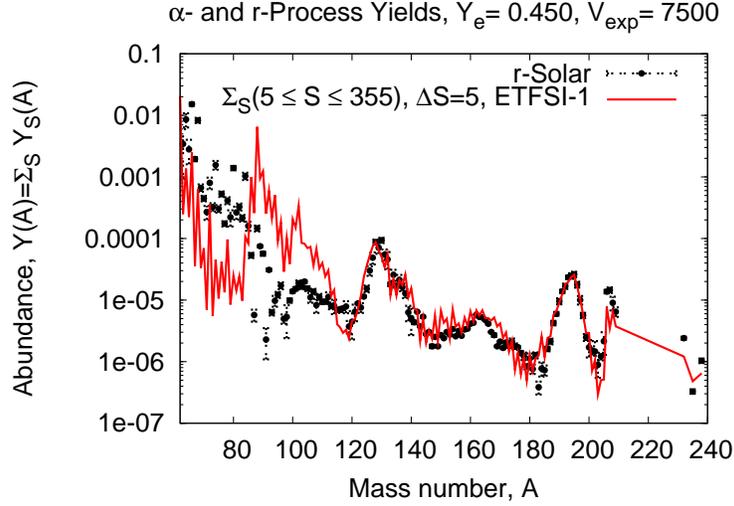,width=100mm,angle=00}}
\caption{Superposition of entropies from $S=5$ up to the maximum entropy $S_{final}=355$ using the unquenchend mass model ETFSI-1. The overall fit of the mass region from A=110 to 238 is excellent.
\label{fig24}}
\end{figure}

\begin{figure}
\centerline{\psfig{file=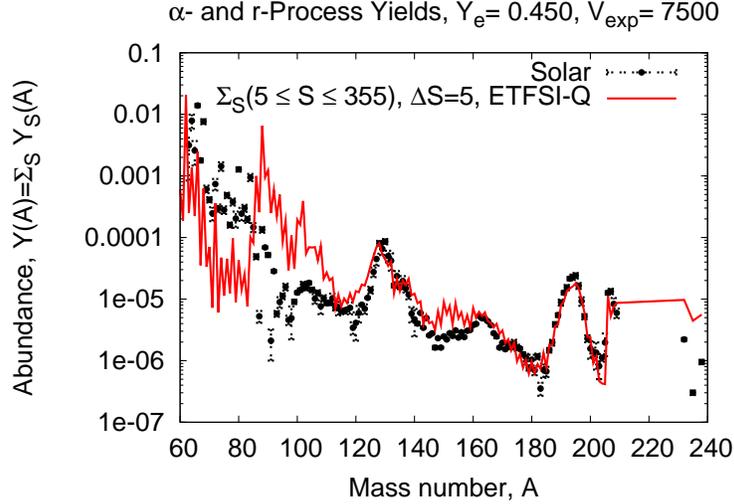,width=100mm,angle=00}}
\caption{Superposition of entropies from $S=5$ up to the maximum entropy $S_{final}=355$ using the quenchend mass model ETFSI-Q. The overall fit of the mass region from A=110 to 238 is good. However, the mass region between A=140 and 160 is somewhat overproduced.
\label{fig25}}
\end{figure}

\begin{figure}
\centerline{\psfig{file=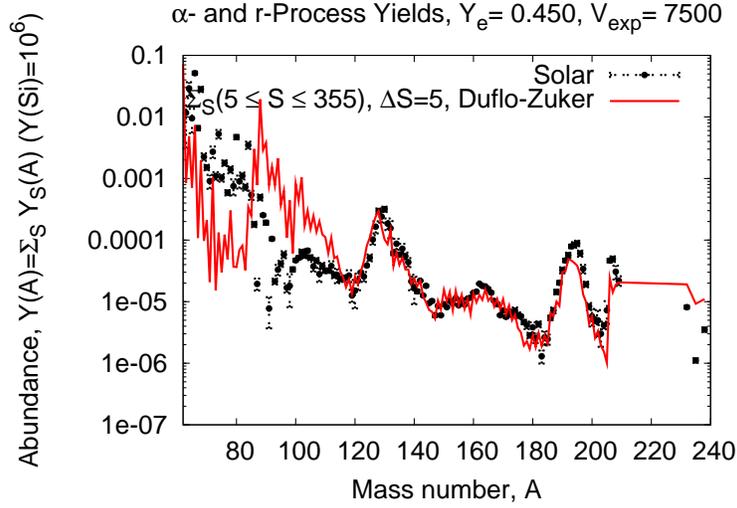,width=100mm,angle=00}}
\caption{Superposition of entropies from $S=5$ up to the maximum entropy $S_{final}=355$ using the mass formula of DUFLO-ZUKER. The overall fit provides the best overall agreement for the REE mass region, but the third r-process peak and the Pb mass region are underproduced. The abundance trough at A=205 is less pronounced than in the case of FRDM.
\label{fig27}}
\end{figure}

\begin{figure}
\centerline{\psfig{file=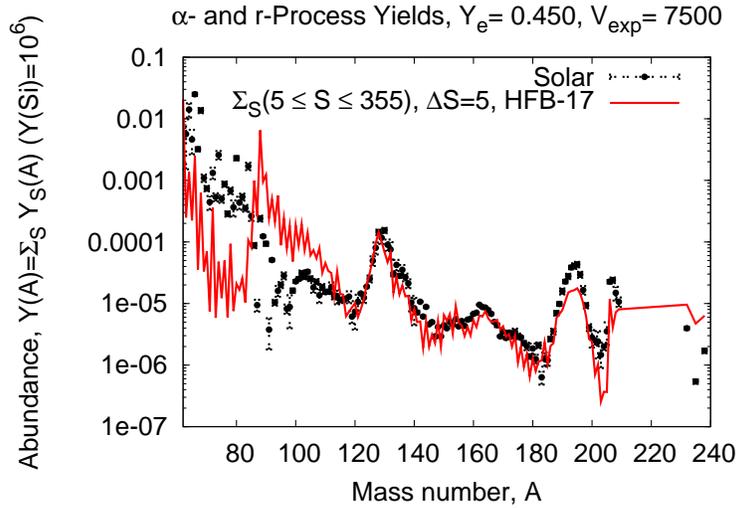,width=100mm,angle=00}}
\caption{Superposition of entropies from $S=5$ up to the maximum entropy $S_{final}=355$ using the mass model HFB-17. Note the slight underabundances in the nuclear shape-transition regions at A$\simeq$145 and A$\simeq$175. The third r-process peak and the Pb mass region are underproduced. The abundance trough at A=205 is more pronounced than in the case of DUFLO-ZUKER.
\label{fig28}}
\end{figure}

$^{130}$Te, $^{162}$Dy and $^{195}$Pt represent the maxima of the A$\simeq$130, the
REE and the A$\simeq$195 r-process abundance peaks. While it is generally
accepted that the two sharp main peaks (both with HWFM widths of about
15 m.u.) are due to the retarded r-matter flow at the N=82 and N=126 shell
closures, the origin of the broad structured REE pygmy peak distribution (with
a total width of about 50 m.u.) as either being a signature of nuclear deformation
or as representing the result of a late abundance enhancement by fission cycling,
still seems to be under debate. In the following, on the basis of Table~\ref{table9} and Figs.~\ref{fig17}-\ref{fig28}, we will discuss in
some detail the formation of the three r-abundance peaks. Table~\ref{table9} shows the isobaric progenitors of the three peak maxima at A=130, 162 and 195, respectively, for the three freeze-out types at different times, temperatures and neutron densities discussed above. Figs.~\ref{fig17}-\ref{fig19} indicate the effects of the late recapture of $\beta$-delayed
neutrons on the final r-abundances.  Figs.~\ref{fig20}-\ref{fig23} show the calculated abundance heights and widths as a function of entropy of the above three peak isotopes for the two selected mass models ETFSI-Q and FRDM. And finally, Figs.~\ref{fig26}-\ref{fig28} present overall fits to the N$_{r,\odot}$ distribution for the five different mass models chosen in the present study. \\
In the historical WP approximation of the r-process at the commonly assumed
freeze-out around T$_9$$\simeq 1$, the main isobaric progenitor of stable $^{130}$Te is the N=82 isotope $^{130}$Cd, for which in the meantime the most important
nuclear-physics properties (T$_{1/2}$, P$_n$ and Q$_\beta$) are experimentally
known \citep{kratz86,hannawald00,dillmann03}. In a fully dynamical r-process
model like our present HEW study, however, the situation is more complicated.
In the following, we will discuss in detail the formation of $^{130}$Cd during the freeze-out phases (see Table~\ref{table9}, and Figs.~\ref{fig11}, \ref{fig14} and \ref{fig17}. For this purpose, we
consider here the (representative) conditions of an entropy of $S=195$, for which
the highest partial $^{130}$Te abundance is obtained.\\
Under these conditions, the inital r-matter flux at a neutron density of $n_n\simeq 10^{27}$
occurs further away from the $\beta$-stability line than in the later freeze-out phases.
As in the classical WP approach, the r-process in the A$\simeq$130 region still
starts to {\it "climb up the N=82 staircase"}~\citep{burb57}, however only up to Z=45 $^{127}$Rh. Already at Z=46 $^{128}$Pd, the r-process begins to break out from the magic
shell, with comparable time-scales for the two competing reactions of further neutron
capture and $\beta$-decay. In the Z=46 isotopic chain, the r-process does produce some
N=82 $^{128}$Pd, but proceeds up to N=84 $^{130}$Pd (with the highest isotopic abundance)
and N=86 $^{132}$Pd. Similarly, also for Z=47 the r-process partly breaks out from
N=82, still producing $^{129}$Ag, but again continuing up to N=84 $^{131}$Ag and N=86 $^{133}$Ag with comparable initial abundances. In this early stage, for the classical Z=48 $^{130}$Cd waiting point, neutron capture out of N=82 still dominates over $\beta$-decay $(\tau_{n\gamma}/ \tau_{\beta}\simeq 0.1)$ thus forming only a small fraction of its total abundance.
Here, the highest initial isotopic yield is obtained for N=86 $^{134}$Cd. Finally, the
classical N=82, Z=49 "break-out" isotope $^{131}$In is passed quickly $(\tau_{n\gamma}/ \tau_{\beta}\simeq 0.02)$, and N=86 $^{135}$In is formed with the highest initial isotopic yield. \\
When focussing in the following on A=130, we recognize that several isotopes lying beyond N=82 with high initial abundances act as $\beta$-decay and $\beta$-delayed 1n to
3n progenitors of neutron magic $^{130}$Cd. As mentioned above, the most important
isobaric r-process nuclide now is N=84 $^{130}$Pd, which contains about $60\%$ of the total
abundances of the early progenitors of $^{130}$Cd. However, due to its high $\beta$-delayed neutron branching ratios (P$_{1n}\simeq 73\%$; P$_{2n}\simeq 18 \%$), and the total P$_{xn}\simeq 24 \%$ of the daughter  $^{130}$Ag, only about $6\%$ of the initial $^{130}$Pd abundance lead to $^{130}$Cd. The next important initial $\beta$-xn progenitors of $^{130}$Cd are $^{132}$Pd and $^{133}$Ag, both with about $20 \%$ progenitor abundance. With these different
decay branches, during the early neutron and chemical freeze-out phases (see Table~\ref{table9}), only a part of the total $^{130}$Cd abundance is produced. Another sizeable fraction comes
from neutron captures of the remaining "free" neutrons. Here again, the most important contribution comes from the initial $^{130}$Pd abundance via its stong $\beta$-delayed 1n decay branch to $^{129}$Ag. When considering the different neutron-capture cross sections for N=82 $^{129}$Ag and its $\beta$-decay daughter N=81 $^{129}$Cd (about a factor 55 in favor of $^{129}$Cd) under the relevant time, temperature and neutron-density conditions at the end of the chemical freeze-out (see Table~\ref{table9}), it is clear that neutron capture predominantly occurs
after $^{129}$Ag $\beta$-decay to $^{129}$Cd into N=82 $^{130}$Cd. The respective production modes of
$^{130}$Cd are reflected by the abundance ratios of Y($^{130}$Cd)/Y($^{129}$Ag) at the beginning
and at the end of the chemical equilibrium. In this freeze-out phase, within about 240 ms,
this ratio changes by about a factor 20 from 0.14 to 2.7. These early abundances are
further modulated during $\beta$-decay back to stability by the capture of the remaining "free"
neutrons and the $\beta$-delayed neutron recapture. The resulting abundance shifts in the A$\simeq$130
peak region from the late recapture of previously emitted $\beta$-delayed neutrons are shown
in Fig.~\ref{fig17}. We see that significant effects occur in the rising wing of the peak. At the top,
the abundance of $^{129}$Xe is reduced by a factor of 2.5, whereas the $^{130}$Te abundance increases by about a factor 2.2.\\
Early debates (see, e.g., \citep{burb57,cam57}) were related to the possible orogin of the REE pygmy peak, either due to mass-asymmetric fission cycling from the trans-actinide region or from the nuclear shapes of the REE progenitor isotopes. In their site-independent approach \citet{kratz93} had already recognized obvious deficiencies of the FRDM mass model in the shape-transition regions before and behind the N=82 and N=126 shell closures, which resulted in an inadequate REE pattern. More detailed network calculations were then performed by \citet{surman97}, again using the FRDM mass model and a simplified set of ``adapted'' $\beta$-decay halfe-lives. They concluded that the REE peak is due to a subtle interplay of nuclear deformation and $\beta$-decay, not occuring in the steady-flow phase of the r-process. At about the same time, the conclusion of the above authors were in principle confirmed by r-abundance calculations of \citet{kratz98}, who compared the REE distributions resulting from (i) the {\it{spherical}} mass model HFB/SkP of \citet{dobaczewski84}, and (ii) the {\it{deformed}} mass model ETFSI-Q of \citet{pearson96}.\\
In the present paper, we use use five selected (deformed) mass models with different microscopic sophistication to reproduce the overall N$_{r,\odot}$ distribution. From Figs.~\ref{fig26}-\ref{fig28} one can clearly see the different success of our dynamical network calcualtions in reproducing the overall shape of the REE pygmy peak. However, with the exception of FRDM, for the other four cases we can consistently confirm and strengthen the earlier conclusions of \citet{surman97} and \citet{kratz98}, that (at least for the HEW scenario) fission cycling seems to be unimportant. And, when considering the FRDM masses, it is highly questionable if its nuclear deficiencies around the magic neutron shells would be ``repaired'' more or less artificially by assuming strong fission cycling at extremely high entropies (e.g., $S>450$ k$_b$/baryon). \\   
From Table~\ref{table9} one can also see that the REE pygmy abundance peak is formed with the ``correct'' shape and relative height in a continuous r-matter flow during feeze-out. It is interesting to note that in the early phase (begin of the chemical equilibrium at about 220 ms) the build-up of the whole REE region is still dominated by neutron captures, forming Z=54 $^{162}$Xe and Z=56 $^{162}$Ba with the highest A=162 isotopic abundances. Within the following about 100 ms, when approaching the end of the chemical equilibrium, the matter flow has continued to the maximum A=162 abundance for Z=58 $^{162}$Ce. Now, up to this mass region $\beta$-decay clearly dominates over further neutron capture. Only in
the late freeze-out phase of the dynamical freeze-out (after about 480 ms), the continued matter flow has formed Z=59 $^{162}$Pr and Z=60 $^{162}$Nd with their maximum yields. Under these low-temperature conditions
$(T_9 \simeq 0.3)$, the density of "free" neutrons has already dropped to $n_n\simeq 10^{17}$ so that by now the whole REE region between
Z=54 and Z=62 is dominated by $\beta$-decays. Finally, as is indicated in
Fig.~\ref{fig18}, during even later times of the decay back to stability, an additional gradual abundance shift to A=162 occurs, which originates from the
recapture of previously emitted $\beta$-delayed neutrons. Thus, the final shape
of the REE pygmy peak is established only at very late non-equilibrium.\\
The formation of the A$\simeq$195 N$_{r,\odot}$ peak, which is related to the
N=126 shell closure, is similar to that of the A$\simeq$130 (N=82) peak but not
exactly the same. There are several differences with respect to both astrophysical as well as nuclear-physics parameters. Starting with a similar initial neutron density of $n_n\simeq 10^{27}$, the higher entropy of $S=270$ (compared to $S=195$ for the A$\simeq$130 peak) leads to a higher neutron to seed abundance ratio of $Y_n/Y_{seed}\simeq 105$ (compared to $Y_n/Y_{seed}=36$ for $S=195$). This results  in an inital r-process path further away from $\beta$-stability than in the A$\simeq$130 region, involving extremely neutron-rich progenitor isotopes with on the average higher $\beta$-delayed neutron branching ratios. In consequence, the description of the formation of the A$\simeq$195 abundance peak and its modulation during the decay back to $\beta$-stability is more complicated. Apart from the "bottle-neck" behavior of the N=126 shell closure, we have to face a detailed interplay between captures of "free"
neutrons, $\beta$-decays, emission of $\beta$-delayed neutrons and their recapture. For some snapshots of the freeze-out during the first 750 ms, see Table~\ref{table9} and Figs.~\ref{fig13}, \ref{fig16} and \ref{fig19}. And at the end, we will argue why the top of the A$\simeq$195 N$_{r,\odot}$ peak occurs with $^{195}$Pt at an odd mass number, and not -- as one would expect -- at an even mass number. \\
In principal similarity to the formation of the early A$\simeq$130 N$_{r,\odot}$ peak, the initial r-matter flux enters the A$\simeq$195 peak region at the
"lighter" N=126 waiting points, producing small fractions of Z=62 $^{188}$Sm to Z=66 $^{192}$Dy. Already from Z=67 on upwards, the early r-process breaks
through the closed shell. In this phase, in the whole mass region neutron capture strongly dominates over $\beta$-decay.\\
At the end of the chemical equilibrium freeze-out phase after about 420 ms (see
Table~\ref{table9}), the abundances of N=126 $^{191}$Tb to $^{196}$Yb have increased between one up to 7 orders of magnitude, all approaching their maximum values during the first 750 ms of the freeze-out. For all "lighter" N=126 isotopes up to Z=68 $^{194}$Er, now $\beta$-decay dominates over neutron capture. Under these conditions, $^{194}$Er has the highest N=126 isotopic abundance. For Z=69 $^{195}$Tm and Z=70 $^{196}$Yb neutron capture and $\beta$-decay time scales become comparable; hence, the break-out from N=126 now has shifted to higher Z values. At the time when the dynamical freeze-out occurs (after about 750 ms; see Table~\ref{table9}), the abundances of the lighter N=126 isotopes up to Z=67 (Ho) have already decreased by more than an order of magnitude due to their $\beta$-decays, whereas under these late conditions finally also the abundances of the "heaviest" N=126 isotopes from Z=71 (Lu) to Z=74 (W) have reached their maximum values. Interestingly, the abundance of Z=68 $^{194}$Er has only decreased by about a factor 4, and the abundance of $^{195}$Tm has remained constant. In this late dynamical freeze-out phase, for all N=126 nuclei $\beta$-decay dominates over neutron capture.\\
As mentioned above, at the end of this section we want to come back to the question why the top of the A$\simeq$195 N$_{r,\odot}$ peak occurs at an odd mass number and not at the - as expected - even mass A=194 of the most abundant classical N=126 waiting-point isotope $^{194}$Er. At the end of the chemical equilibrium and in the subsequent dynamical freeze-out phase, a subtle interplay of several $\beta$-decay, $\beta$1n- to $\beta$3n-decay and neutron-capture channels occur, which lead to kind of reaction cycles. Recapture of $\beta$-delayed neutrons does not yet play a significant role. Under these conditions, the maximum abundance in the peak is still $^{194}$Er closely followed by $^{195}$Tm. When neglecting $\beta$-delayed neutron recapture, obviously this signature is held during the whole decay back to stability, resulting in a final abundance ratio of Y($^{194}$Pt)/Y($^{195}$Pt)$\simeq$1.3. However, when allowing the recapture of previously emitted $\beta$-delayed neutrons, which
occurs relatively late during the $\beta$-backdecay, a significant upwards abundance shift with Z occurs in the whole rising wing up to the top of the peak. Beyond A=195, for a few masses the reverse effect is observed. This is shown in Fig.~\ref{fig19}. In our calculations, the final abundance ratio now changes to Y($^{194}$Pt)/Y($^{195}$Pt)$\simeq$0.5. Hence, we again conclude that the recapture of $\beta$-delayed neutrons has to be included in detailed r-process calculations.                         
\section{REPRODUCTION OF THE SOLAR-SYSTEM R-PROCESS ABUNDANCES}
\subsection{The Entropy Weights}
In our attempt to not only understand the behavior of individual 
entropy components, but rather to find an explanation of the solar system isotopic r-abundance residuals $\mbox{N}_{r,\odot}\simeq\mbox{N}_{\odot}-N_{s,\odot}$, we need to consider a recipe for the weights in such entropy superpositions, similar to
superposition of neutron number densities in classical r-process calculations \citep{kratz93,cowan99,pfeiffer01b,kratz07}. Hereafter, we will discuss two simple approaches, assuming the ejection of neutrino-wind elements per equidistant entropy interval with equal masses $(dM(S)/dS=const)$  or with equal volume $(dV(S)/dS=const)$, where the entropy spans from $S=5$ to $S_{final}$. This is a pure assumption not resulting from from dynamic explosion calculations; but we will show that this simple weighting leads to excellent fits to the solar r-process abundances between the rising wing of the A$\simeq$130 peak and the Pb-Bi peak.
\subsubsection{Ejecta per equidistant entropy with equal volumes}
If the ejected neutrino-wind elements are equal-sized (hereafter, assumption 1), one obtains
$$\frac{M(S)}{\rho(S)}=const.$$ Since $S\sim 1/\rho$, one obtains for an entropy $S$
\begin{equation}\label{weight3}
M(S)\, S=M(S_{ref})\, S_{ref}=const,
\end{equation}
where $S_{ref}$ represents an arbitrary reference entropy which we set equal to 5 k$_{B}$/baryon, the smallest entropy in our network calulations. From Eq.~\ref{weight3}, one infers that the entropy weight is
\begin{equation}\label{weight1}
w(S)=\frac{M(S)}{M(S_{ref})}=\frac{S_{ref}}{S}.
\end{equation}
Hence, by integrating over different entropies, the accumulated abundance $Y_{sum}(Z,A)$ of a given nucleus $(Z,A)$ is
\begin{equation}\label{weight1_1}
Y_{sum}(Z,A)=\sum_{S=5}^{S_{final}} \frac{S_{ref}}{S} Y_{S}(Z,A).
\end{equation}
Therefore, the ejected mass fraction of heavy material per entropy is given by $$X_{heavy}(S)=w(S)(1-X_{\alpha}).$$ Figs.~\ref{fig20} and~\ref{fig21} show the calulated abundances of $^{130}$Te, $^{162}$Dy and $^{195}$Pt as a function of entropy. The agreement of the peak heights of our simulations with the corresponding solar-system r-abundances is excellent for the mass models ETFSI-Q and FRDM. However, the latter mass model underproduces the REE mass region because of its very strong N=82-shell closure.
\subsubsection{Ejecta per equidistant entropy with equal masses}
Here we show (as a different example) superpositions with equal amounts of ejected mass per entropy interval (hereafter, assumption 2). When the ejected masses per entropy are equal, the entropy weights will be the same, i.e., $w(S)=1$. The accumulated abundance of a given nucleus then is
\begin{equation}\label{weight2}
Y_{sum}(Z,A)=\sum_{S=5}^{S_{final}} Y_{S}(Z,A).
\end{equation}
Figs.~\ref{fig22} and~\ref{fig23} show the results for ETFSI-Q and FRDM. Both mass models show similar (but slightly worse) fits to the solar r-abundances and thus indicate that within the expected uncertainties both assumptions 1 and 2 yield good agreement. This might indicate what one would expect from realistic scenarios. For the attempt to reproduce the solar r-abundances, in the following we use the assumption 1.
\subsection{Fitting the Solar r-Process Abundances}
\begin{figure}
\centerline{\psfig{file=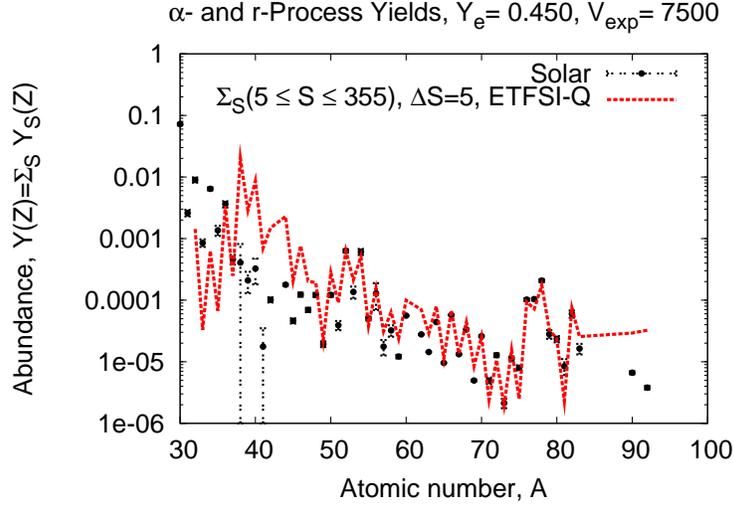,width=100mm,angle=00}}
\caption{Solar-system elemental abundances obtained with the same method discribed in Fig.~\ref{fig24}. Significant deviations of the fit from the solar r-abundances occur at the elements below Cd (Z$<$48). For further discussion, see text.
\label{fig29}}
\end{figure}

\begin{figure}
\centerline{\psfig{file=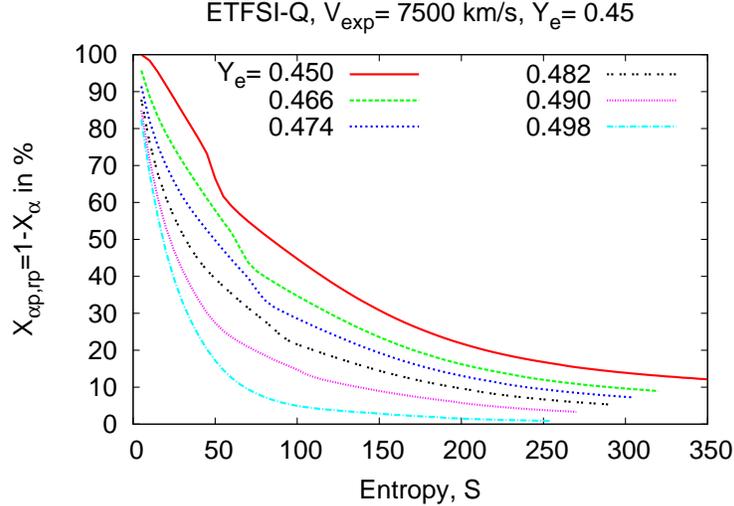,width=100mm,angle=00}}
\caption{Mass fractions of the expected heavy ejecta (A$>$4) as a function of entropy $S$ and electron abundance $Y_e$ in case of ejection of neutrino-wind elements per equidistant entropy interval. The mass fraction of the heavy ejecta decreases with increasing $S$ and $Y_e$.  
\label{fig30}}
\end{figure}

\begin{figure}
\centerline{\psfig{file=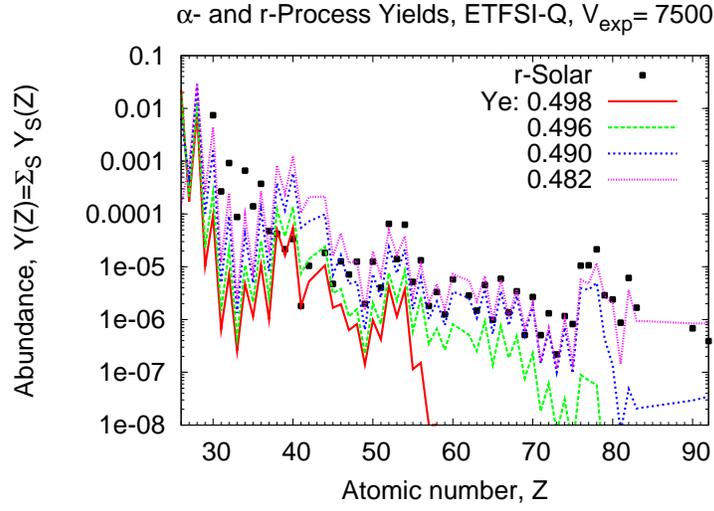,width=100mm,angle=00}}
\caption{Solar-system elemental yields are compared to the results of 4 electron abundances ranging from 0.498 down to 0.482. For each $Y_e$ the superposition of the entropies spans from $S=5$ to the maximum entropy $S_{final}(Y_e)$. The elemental solar r-abundances are scaled such that $Y_{r,\odot}(\mbox{Nb})=Y_{0.498}\mbox{(Nb})$. The element Nb is chosen as reference because it has only one stable isotope and therefore can also be used as reference for the isotopic solar r-abundances.
\label{fig31}}
\end{figure}

In the past, many attempts were undertaken to reproduce the solar r-process abundances. In the frame of the WP approximation, based on an $(n,\gamma)-(\gamma,n)$-equilibrium with static temperatures and neutron number densities, it has been shown that only a superposition of a several of neutron number density components with weights given by the realistic heights of the three N$_{r,\odot}$-peaks (Y($^{80}$Se)$\simeq 24.3$, Y($^{130}$Te)$\simeq 1.69$, Y($^{195}$Pt)$\simeq 0.457$ \citep{kaeppeler89}) can achieve this goal \citep{kratz93}.
~\citet{freiburghaus99} used an entropy-based superposition with an analytical fit function $g(S)=x_1e^{-x_2S}$, where $x_1$ and $x_2$ represent two fit parameters. We follow the method discribed in this work by using our above assumption 1. Figs.~\ref{fig26} to \ref{fig28} show the results for a selected constant expansion velocity of $V_{exp}=7500$ km/s and a specific electron abundance $Y_e=0.45$. It is surprising that this simple parametrisation of the HEW components provides a good fit to the overall solar r-abundances above A=110, especially with the mass models ETFSI-1 and ETFSI-Q. The mass formula of DUFLO-ZUKER slightly underproduces the A$\simeq$195 and the Pb mass regions, and the new mass model HFB-17 shows deficiencies in the A$\simeq$145 and A$\simeq$175 mass regions where nuclear phase transitions occur. In these regions, the mass model prescriptions may still have systematic deficiencies. HFB-17 also underproduces the A$\simeq$195 and the Pb mass regions. The mass model FRDM has defeciencies in producing the REE, the A$\simeq$200 and the Pb mass regions.\\
The over-production of the mass region below A$<110$ by the HEW is well-known and was discussed by several authors in the past (see, e.g., \citep{woosley94,freiburghaus99}). It has been referred to as a model deficiency due to the $\alpha$-rich freeze-out component, where the r-process cannot start because the neutron to seed ratio is still smaller than unity. Fig.~\ref{fig29} shows the fit to the solar elemental r-abundances by using the mass model ETFSI-Q. It is quite evident from the this figure that the abundance over-production mentioned above concerns the elements between Sr and Ag. However, in our new understanding of the HEW nucleosynthesis one can avoid this disagreement by considering an additional superposition in terms of $Y_e$.\\
Since $S$ increases and $Y_e$ decreases with time, it is reasonable to consider the ejecta of a core-collapse supernova explosion as a mixture of different $S$ and $Y_e$-components. Fig.~\ref{fig30} shows the mass fraction of the expected heavy ejecta (A$>4$) for a selection of $Y_e$-values as a function of $S$. Thereby, one can see that the weight of the ejecta increases as $Y_e$ decreases. Fig~\ref{fig31} shows that the whole mass region from Sr (Z=38) up to U (Z=92) can be fitted by using 4 different $Y_e$-values (0.498, 0,496, 0,490 and 0.482). Thereby, we have normalized the solar r-abundances so that $Y(\mbox{Nb})_{r,\odot}=Y(\mbox{Nb})_{Y_{e}=0.498}$. We have chosen the element Nb because it can serve as normalization for the elemental and isotopic solar r-abundances since it has only one stable isotope. From Fig.~\ref{fig31} one can infer:
\begin{itemize}
\item Sr, Y, Zr and Nb (Z=38-41) are formed under condition of $Y_e=0.498$
\item Mo and Ru (Z=42 and 44) are formed under conditions of $Y_e=0.496$
\item Rh, Pd and Ag (Z=45-47) are formed under further decreasing $Y_e$-conditions of $Y_e=0.490$
\item Cd and the elements beyond (Z$\ge 48$) are formed under more neutron-rich conditions of $Y_e=0.482$
\end{itemize}
Thus, the well-known over-production of the mass region from A$=88$ to 110 in
the HEW in Figs.~\ref{fig26} to \ref{fig28}, is due to the fact that only one
$Y_e$ is used. A superposition of several $Y_e$-values (as e.g., performed by \citet{meyer92} for a selected entropy), however in addition to the superposition of $S$-values used in our study, could help to resolve this problem.\\
\subsection{The Amount of the r-Process Ejecta}
Based on our assumptions concerning the entropy weights, which provide a good fit 
to the solar r-abundances beyond the A=110 mass region, we will estimate the 
amount of the (neutron-capture) r-process material which is ejected,    
\begin{equation}\label{ejecta1}
 \mbox{M}_{r}\approx \frac{4}{3}\, \pi\, R_{0}^{3}(t=0)\, \int_{S_{initial}}^{S_{final}} dS (1-4Y_{\alpha})\rho(S,t=0).
\end{equation}
The range from $S_{initial}$ to $S_{final}$ indicates the entropy interval 
which permits the production of r-process nuclei. Furthermore, our calculations
show that the entropy-dependent r-process mass fraction $ X_{r}=1-4Y_{\alpha}$ 
can be well approximated by a function $f(S)=x_{1}S^{x_{2}}$. Eq.~\ref{ejecta1} then reads

 \begin{equation}\label{ejecta2}
 \mbox{M}_{r}\approx \frac{4}{3}\, \pi\, R_{0}^{3}(t=0)x_{1}\int dS\; S^{x_{2}}\rho(S,t=0).
\end{equation} 
Replacing the term of the matter density in Eq.~\ref{ejecta2} by its term in Eq.~\ref{dichte}, one obtains

 \begin{equation}\label{ejecta3}
 \mbox{M}_{r}\approx 1.1\, 10^{-3} M_{\odot}\, x_{1}\int dS\; \frac{S^{x_{2}}}{S},
\end{equation} 
in units of solar masses $M_{\odot}$. The paramters $x_{1}$ and $x_{2}$ vary with $V_{exp}$ and $Y_{e}$. If we complete the integration in Eq.~\ref{ejecta3}, one finally gets
 \begin{equation}\label{ejecta4}
 \mbox{M}_{r}\approx 1.1\, 10^{-3} M_{\odot}\, \frac{x_{1}}{x_{2}}\left(S_{final}^{x_{2}}-S_{initial}^{x_{2}}\right).
\end{equation} 
Tables~\ref{table10} shows the results according to Eq.~\ref{ejecta4} for two expansion velocities. The amount of the estimated mass of the ejecta then depends only on the electron abundance $Y_{e}$. For instance, at a given expansion speed a $Y_{e}= 0.45$ will produce roughly 10 times more r-process material than 0.49. This is due to the higher entropy needed to start an r-process with the latter $Y_{e}$-value.   

\section{SUMMARY AND CONCLUSIONS}
We have performed large-scale parameterized dynamical network calculations in
the context of an adiabatically expanding high-entropy wind as expected in
core-collapse supernovae. We have used four mass models and a mass formula to test in great detail the validity of the $(n,\gamma)-(\gamma,n)$- and the
$\beta$-flow equilibria and the time intervals during which they are fulfilled during an r-process. We have defined different kinds of freeze-out, (a) chemical, (b) neutron and (c) dynamical, that are related to (a) the break-out of the $(n,\gamma)-(\gamma,n)$-equilibrium, (b) to the consumption of neutrons, i.e., $Y_n/Y_{heavy}<1$, and (c) to the time at which the neutron depletion timescale becomes larger than the hydrodynamical timescale, i.e., $\left|Y_n/ \dot{Y}_n\right|>\left|\rho(t)/\dot{\rho}(t)\right|$, respectively. In our attempt to provide the best possible fit to the solar r-process abundances, we have explored the consequences of two simple assumptions concerning the matter ejection from the newly born proto-neutron star in the HEW. The first assumption is based on ejection of neutrino-wind elements with equal sizes, and the second assumption is based on the ejection of neutrino-wind elements with equal masses per equidistant entropy interval. Since the first assumption yielded exactly the same ratios of the solar r-abundance peaks ($^{130}$Te$_{r,\odot}$/$^{162}$Dy$_{r,\odot}\approx 16$ and $^{130}$Te$_{r,\odot}$/$^{195}$Pt$_{r,\odot}\approx 3.5$) we made use of it to determine the entropy weights, in order to superpose the entropy yields and to estimate the amount of the r-process material which can be ejected by the HEW of a core-collapse supernova explosion.\\
Furthermore, we can draw the following conclusions:
\begin{itemize}
\item The entropies generated by the HEW exhibit maximum values after which the production of heavy seed nuclei via the bottle-neck nuclear reactions (Eqs.~\ref{triple} and~\ref{aan}) fails. We have labeled those entropies with $S_{final}$. The seed distributions beyond $S_{final}$, consisting mainly of $\alpha$-particles, neutrons, protons and traces of $^{12}$C and $^{16}$O, do not allow an r-process to proceed because of the bottle-neck regions at $A=5$ and 8. Furthermore,  for a given expansion velocity of the matter in the HEW, the maximum entropies $S_{final}$ increase with decreasing $Y_e$ because the three-body bottle-neck nuclear reaction (Eq.~\ref{aan}) depends on the availability of neutrons in the wind. It can therefore proceed at lower densities than the triple-$\alpha$-reaction since an uncharged particle is involved. An interesting observation is that (independent of $S_{final}$) $Y_e$-values larger than 0.49 do not yield neutron to seed ratios large enough to synthesize the third r-process peak elements and Th and U.
\item The mass region between Fe and Zr (depending somewhat on $Y_e$) which is
historically thought to be the beginning of the "weak" r-process, is obviously
formed by rapid charged-particle reactions in the HEW at low entropies. From
about Nb on, our model predicts a smooth transition into a true rapid neutron-capture
r-process. This seems, indeed, to be confirmed by recent astrophysical and
cosmochemical observations (see, e.g., \citep{kratz08,farouqi09a,farouqi09b}). Therefore, in this mass region the classical definition of the so-called
Solar System (SS) r-process "residuals" as $\mbox{N}{_\odot}-\mbox{N}_{s,\odot}=\mbox{N}_{r,\odot}$ should no longer be applied.
\item While in our calculations the lightest trans-Fe elements up to about Kr are underproduced
by one to two orders of magnitude relative to the SS abundances, the region between
Sr and Ag appears to be overproduced by about a factor 4 when using the full entropy
range. These two kinds of differences relative to N$_{r,\odot}$ may be explained in the
following way: (i) for the region between Fe and about Kr by missing abundance contributions
from the $\nu$-p process (see, e.g., \citep{frohlich06a,frohlich06b}) and/or by additional contributions from a
new type of rs-process \citep{pignatari08}; and (ii) for the region between Sr and Ag by the
assumption of a unique, constant electron abundance in the HEW when integrating over
the entropy yields. 
\item  We also have shown by our HEW r-process studies that for electron abundances slightly
below $Y_e = 0.50$, e.g., for $Y_e = 0.498$, only the mass region below the A$\simeq$130
peak can be formed. The region of the classical "main" r-process up to the full 3rd peak
requires somewhat more neutron-rich winds with $Y_e$ values in the range 0.48; and finally,
a full r-process including sufficient abundances of the actinides Th and U can only be
produced with even lower $Y_e$ values in the range below about 0.478. This is due to the
fact that the amount of heavy ejecta as a function of $S$ decreases with increasing $Y_e$. 
\item The SS r-process residuals beyond A$\simeq$110 can be well fitted by assuming
that the HEW of core-collapse supernova explosions will eject a series of neutrino wind elements per equidistant entropy interval. Thereby, the relative heights of the A$\simeq$130
and A$\simeq$195 major peaks as well as the intermediate REE pygmy peak in the
r-abundance fits reproduce the N$_{r,\odot}$ distributions quite well for a variety of
microscopic global mass models. Depending on $Y_e$, the total r-process masses lie
between between $10^{-6}$ and $10^{-4}$ M$_{\odot}$, in good agreement with Galactic chemical evolution studies~\citep{truran71,hillebrandt78,ctt91}. Both, nuclear-physics and astrophysical parameters have sizeable effects on the
process duration to overcome the major bottle neck in the r-process matter flow at
the N=82 shell thereby forming the A$\simeq$130 abundance peak, as well as the
total r-process duration to produce sufficient amounts of Th and U. For example, by
using the same astrophysical parameters ($Y_e=0.45$, $S\le280$ and $V_{exp}=7500$ km/s)
for the quenched mass model ETFSI-Q the full r-process up to the actinides is about
a factor three "faster" than for the older, unquenched mass model FRDM. Therefore,
for any type of modern r-process calculation in whatever astropyhsical scenario, a
sophisticated nuclear-physics input, in particular around the
magic neutron shells, remains crucial. 
\item The morphology of the r-process in the HEW can be depicted as:
\begin{enumerate}
\item A "hot" r-process (in full chemical equilibrium), which produces the mass region up to the A$\simeq$130 r-process peak;
  \item a "hybrid" r-process (in partial chemical equilibrium), which is responsible for
      producing the REE region; and
  \item a "cold" r-process (out of chemical equilibrium), which forms the A$\simeq$195
      peak and the actinides. 
      \end{enumerate}
\item The REE pygmy peak of the SS r-process abundances in between the two major peaks
  at A$\simeq$130 (caused by the N=82 shell closure) and at A$\simeq$195 (caused
  by the N=126 magic shell) is formed in rather small entropy ranges for the more recent
  microscopic mass models, with the exception of the still frequently used older macroscopic-microscopic model FRDM. With this, our results clearly show that with an appropriate
  choice of the nuclear-physics input, fission cycling is not neceessary to reproduce the
  correct shape of the REE pattern. 
\item Without considering effects from the emission and recapture of $\beta$-delayed neutrons, the
  two main r-abundance peaks with their tops at A=130 and A=195 would occur at A=129
  and A=194. The late recapture of $\beta$-delayed neutrons is the main reason for the gradual
  shifts from A=129 to the classical major N=82 "waiting-point" isotope $^{130}$Cd in a
  quasi-chemical equilibrium, and the respective shift from A=194 to the classical N=126
  "waiting-point" nuclide $^{195}$Tm under non-equilibrium conditions. 
\item The $\beta$-flow equilibrium is confirmed not to occur globally during the whole r-process. Only
  local equilibria are achieved, partly at the magic neutron shells and in between the different shell closures. 
\end{itemize} 

\section{Acknowledgments}
This work is supported in part at the University of Chicago by
the National Science Foundation under Grants PHY 02-16783
and PHY 08-22648 for the Physics Frontier Center "Joint Institute for 
Nuclear Astrophysics" (JINA), by the "Deutsche Forschungsgemeinschaft" (DFG) under contract KR 806/13-1, by the Helmholtz Gemeienschaft under grant VH-VI-061 and by the Swiss National Science Foundation (SNF).\\
JWT and KF acknowledge support from the Argonne National Laboratory, operated under contract No. DE-AC02-06CH11357 with the DOE.\\
We thank an anonymous referee for his careful reading of the manuscript,
the helpful suggestions for corrections which led to an improved clarity of our results.


\clearpage
\clearpage
\clearpage
\clearpage
\clearpage
\clearpage
\clearpage
\clearpage
\clearpage
\clearpage
\clearpage
\clearpage
\clearpage

\clearpage
\clearpage
\clearpage
\begin{table}
\begin{center}
\caption{\label{table1}Dimension of the full reaction network used until 
charged-particle freeze-out. The particle unstable isotopes $^{5}{\mbox{He}}$, $^{8}{\mbox{Be}}$ and $^{9}{\mbox{B}}$ are 
  included implicitely as intermediate nuclei in (formal) two-body representations of three-body reactions. The network contains 1989 nuclei and the neutron.}
\vspace{10mm}
\begin{tabular}{l l c r | l l c r}
\hline
Element& Z &A$_{\mbox{min}}$&A$_{\mbox{max}}$ & Element& Z &A$_{\mbox{min}}$& A$_{\mbox{max}}$\\ \hline
H &1& 1 & 3      &Cr &24  &38 &86 \\
He& 2& 3 & 6     &Mn &25  &40 &89 \\
Li&3 & 6 & 9     &Fe&26   &42 &92\\
Be &4   &7   &12 &Co &27  &44 &96\\
B  &5   &8   &14 &Ni &28  &46 &99\\
C  &6   &9   &18 &Cu &29  &48 &102 \\
N  &7   &11  &21 &Zn &30  &51 &105 \\
0  &8   &13  &22 &Ga &31  &53 &108 \\
F  &9   &14  &26 &Ge &32  &55 &112 \\
Ne &10  &15  &41 &As &33  &57 &115 \\
Na &11  &17  &44 &Se &34  &59 &118 \\
Mg &12  &19  &47 &Br &35  &61 &121 \\
Al &13  &21  &51 &Kr &36  &63 &124 \\
Si &14  &22  &54 &Rb &37  &66 &128 \\
P  &15  &23  &57 &Sr &38  &68 &131 \\
S  &16  &24  &60 &Y  &39  &70 &134 \\
Cl &17  &26  &63 &Zr &40  &72 &137 \\
Ar &18  &27  &67 &Nb &41  &74 &140 \\
K  &19  &29  &70 &Mo &42  &77 &144 \\
Ca &20  &30  &73 &Tc &43  &79 &147 \\
Sc &21  &32  &76 &Ru &44  &81 &150 \\
Ti &22  &34  &80 &Rh &45  &83 &153 \\
V  &23  &36  &83 &Pd &46  &86 &156 \\
\hline
\end{tabular}
\end{center}
\end{table}

\begin{table}
\caption{\label{table2} The mass fractions of $\alpha$-particles ($X_{max}$) and neutrons ($X_{n}$) out of which the seed nuclei are formed.}
\begin{center}
\begin{tabular}{c|c|c}
\hline
\hline
 $Y_{e}$  &$X_{max}$ in $\%$  & $X_{n}$ in $\%$\\   
\hline
\hline
 0.499 & 99.8 & 0.2\\
 0.498 & 99.6 & 0.4\\
 0.496 & 99.2 & 0.8\\
 0.49 & 98 & 2\\
 0.48& 96 & 4\\
0.47& 94 & 6\\
0.46& 92 & 8\\
0.45& 90 & 10\\
0.40& 80 & 20\\
\hline
\hline
\end{tabular}
\end{center}
\end{table}
\begin{table}
\caption{\label{table3} Abundances and mass fractions obtained for a boundary entropy beyond $S_{final}$. The model parameters are $S= 278$, $Y_{e}= 0.49$ and $V_{exp}= 7500$ km/s $(\tau_{exp}= 35$ ms). For further details, see text.}
\begin{center}
\begin{tabular}{c|c|c}
\hline
\hline
Baryon & Abundance (Y)&  Mass fraction (X in $\%$)\\   
\hline
\hline
 $^{4}\mbox{He}$ & 0.2439 & 97.55\\
              n & 0.22 $10^{-1}$&2.2  \\
              H & 0.23 $10^{-2}$&0.23  \\
 $^{2}\mbox{H}$ & 0.28 $10^{-6}$&0.56 $10^{-6}$  \\
 $^{3}\mbox{H}$ & 0.73 $10^{-7}$&0.22 $10^{-6}$  \\
 $^{12}\mbox{C}$ & 0.22 $10^{-8}$&0.26 $10^{-7}$  \\
 $^{16}\mbox{O}$ & 0.11 $10^{-8}$&0.18 $10^{-7}$  \\
 $^{3}\mbox{He}$ & 0.41 $10^{-9}$&0.12 $10^{-8}$  \\
\hline
\hline
\end{tabular}
\end{center}
\end{table}
\newpage
\begin{table}
\caption{\label{table4} The $\alpha$-rich freeze-out conditions for an r-process at two expansion velocities $V_{exp}= 7500$ and 15000 km/s which correspond to expansion timescales of 35 and 15 ms, respectively. For a given $Y_e$, $S_{initial}$ is the entropy which initiates an r-rpocess $(Y_n/Y_{seed}\approx 1)$, $S_{final}$ is the last entropy which is able to build heavy seed nuclei (beyond it no r-process can take place), $\Delta S=S_{final}-S_{initial}$, and $\left(\frac{Y_{n}}{Y_{seed}}\right)_{final}$ represents the maximum $Y_n/Y_{seed}$ ratio which can be attained when $S=S_{final}$. For further details, see discussion in text.}
\begin{center}
\begin{tabular}{|c |c c c c |c c c c|}
\hline
 $Y_e$&\multicolumn{4}{|c}{7500 km/s}&\multicolumn{4}{|c|}{15000 km/s}\\
\hline
 & $S_{initial}$ &$S_{final}$ &$\Delta S$&$\left(\frac{Y_{n}}{Y_{seed}}\right)_{final}$& $S_{initial}$ &$S_{final}$ &$\Delta S$&$\left(\frac{Y_{n}}{Y_{seed}}\right)_{final}$ \\
\hline
0.499&   206&   253&    47&  17&   156&   193&    37&    17\\
0.498&   185&   258&    73&  37&   143&   194&    51&      32\\
0.496&   162&   263&   101&  60&   128&   198&    70&       50\\
0.494&   153&   267&   114&  72&   121&   202&    81&      62\\
0.492&   147&   272&   125&  84&   117&   205&    88&      68\\
0.490&   142&   277&   135&  96&   113&   209&    96&      78\\
0.486&   136&   283&   147&  109&   109&   216&   107&      92\\
0.482&   131&   295&   164&  132&   105&   223&   118&      107\\
0.478&   126&   303&   177&  148&   101&   229&   128&   120\\
0.474&   122&   311&   189&  167&    98&   236&   138&      139\\
0.470&   118&   319&   201&  188&    95&   242&   147&      155\\
0.466&   114&   327&   213&  212&    92&   248&   156&      174\\
0.462&   111&   334&   223&  233&    89&   254&   165&      195\\
0.458&   107&   341&   234&  258&    87&   260&   173&      219\\
0.454&   104&   348&   244&  286&    84&   265&   181&      239\\
0.450&   100&   354&   254&  311&    81&   270&   189&      261\\
0.400&    53&   416&   363&  755&    43&   320&   277&      655\\
\hline
\end{tabular}
\end{center}
\end{table}

\begin{table}
\caption{\label{table8} r-Process simulations with $V_{exp}= 7500$ km/s and $Y_{e}= 0.45$. The first column shows a choice of entropies, the second shows the corresponding r-process strength, i.e., the $Y_n/Y_{seed}$ ratio. The third and sixth columns represent the r-process durations, i.e., the time between the freeze-out of the charged-particle reactions and the freeze-out of the following r-process. The fourth and seventh columns show the r-process freeze-out temperatures, and the fifth and last columns show the nucleus with the maximum abundance after decay back to stability for the FRDM and ETFSI-Q mass models, respectively. Note the earlier onset of fission cycling with FRDM.}
\begin{center}
\begin{tabular}{|c|c|l l l |l l l|}
\hline
 Entropy&Strength &\multicolumn{3}{|c}{FRDM}&\multicolumn{3}{|c|}{ETFSI-Q}\\
\hline
 $S$ (k$_{B}$/baryon) &$Y_n/Y_{seed}$      &$t_{freeze}$ (ms) &$T_{9}$&Maximum &$t_{freeze}$ (ms) &$T_{9}$&Maximum \\
\hline

    100& 1 &        1&  2.95&   $^{92}\mbox{Zr}$&           1&  2.93&    $^{92}\mbox{Zr}$\\
    130& 6 &        21&  2.14&   $^{104}\mbox{Ru}$&           34&  1.81&    $^{102}\mbox{Ru}$\\
    175&23   &       91&  1.09&   $^{126}\mbox{Te}$&          86&  1.13&    $^{128}\mbox{Te}$\\
    195&   34  &      197&  0.626&   $^{129}\mbox{Xe}$&        138&  0.818&    $^{130}\mbox{Te}$\\
    215&48   &      327&  0.411&   $^{132}\mbox{Xe}$&         208&  0.600&    $^{140}\mbox{Ce}$\\
    230& 61   &      375&  0.365&   $^{132}\mbox{Xe}$&        241&  0.531&    $^{154}\mbox{Sm}$\\
    236&   66 &     398&  0.346&   $^{132}\mbox{Xe}$&         253&  0.511&    $^{162}\mbox{Dy}$\\
    260&92    &     528&  0.269&   $^{195}\mbox{Pt}$&        322&  0.417&    $^{192}\mbox{ Os}$\\
    270&  107  &    600&  0.239&   $^{196}\mbox{Pt}$&       398&  0.347&    $^{195}\mbox{Pt}$\\
    280&   121 &    684&  0.212&   $^{129}\mbox{Xe}$&        470&  0.299&    $^{206}\mbox{Pb}$\\
    295&146    &     860&  0.171&   $^{129}\mbox{Xe}$&      586&  0.244&    $^{232}\mbox{Th}$\\
    300&154     &    926&  0.159&   $^{129}\mbox{Xe}$&       629&  0.229&    $^{130}\mbox{Te}$\\
\hline
\hline
\end{tabular}
\end{center}
\end{table}

\begin{table}
\caption{\label{table9} The main isobaric progenitors of $^{130}$Te, $^{162}$Dy and $^{195}$Pt using the mass model ETFSI-Q. The model parameters are $Y_{e}= 0.45$, $V_{exp}= 7500$ km/s, and $S=195$, 236 and 270. The neutron, chemical and dynamical freeze-outs mean $Y_n$/$Y_{heavy}<1$, the break-out of the $(n,\gamma)-(\gamma,n)$-equilibrium and $|Y_n(t)$/$\dot{Y}_n(t)|>|\rho(t)/\dot{\rho}(t)|$, respectively. The superscripts $b$ and $e$ denote the ``begin'' and ``end'' of the $(n,\gamma)-(\gamma,n)$-equilibrium, respectively.}
\begin{center}
\begin{tabular}{l|l|c|c|l|l|l|c}
\hline
\hline

Peak &Freeze-out& Progenitor&Y $(10^{-5})$& Time (ms)& T$_9$&$n_n$ (cm$^{-3}$)&$Y_n$/$Y_{heavy}$\\   
\hline
$^{130}$Te &Neutron        &$^{130}$Pd &50 &138& 0.82&$10^{23}$&1  \\
           &Chemical$^b$  &$^{130}$Pd &23&171& 0.71&$10^{21}$&$10^{-2}$ \\
           &Chemical$^e$  &$^{130}$Cd &21&414& 0.33&$10^{18}$&$10^{-4}$ \\
           &Dynamical     &$^{130}$Cd &16&415& 0.33&$10^{18}$&$ 10^{-4}$ \\
\hline
$^{162}$Dy &Chemical$^b$ &$^{162}$Xe&1.3&220& 0.57&$ 10^{24}$&9 \\
           &Neutron     &$^{162}$Ba&4.0&252& 0.52&$ 10^{23}$&1 \\
           &Chemical$^e$&$^{162}$Ce&1.9&331& 0.41&$ 10^{19}$&$ 10^{-4}$\\
           &Dynamical   &$^{162}$Nd&0.7&479& 0.29&$ 10^{17}$&$ 10^{-5}$ \\
\hline
$^{195}$Pt &Chemical$^b$  &$^{195}$Ho&1.9 $10^{-3}$ &212& 0.59&$ 10^{24}$&47 \\
           &Neutron      &$^{195}$Tm&1.9  &396& 0.35&$ 10^{22}$&1 \\
           &Chemical$^e$ &$^{195}$Tm&4.3  &422& 0.33&$ 10^{21}$&0.2 \\
           &Dynamical    &$^{195}$Yb&12  &701& 0.21&$ 10^{18}$&$10^{-4}$ \\
\hline
\hline
\end{tabular}
\end{center}
\end{table}

\begin{table}
\caption{\label{table10} The total mass of r-process material which can be ejected from the HEW of core-collapse supernovae for two different expansion velocities of $V_{exp}=7500$ and 15000 km/s. For each electron abundance $Y_e$, the integration over the entropies starts from $S_{initial}$ $(Y_n/Y_{seed}\approx 1$) to the maximum entropy $S_{final}(Y_e)$. For more details, see text.}
\begin{center}
\begin{tabular}{c}
\hline
\hline
         Mass in $10^{-4}\mbox{M}_{\odot}$\\   
\hline
\hline
\begin{tabular}{c|c|c}
 $Y_{e}$  &7500 km/s & 15000 km/s\\ 
\hline
\hline

0.450     &      3.42  &      3.30  \\
0.458     &      2.71  &      2.57   \\
0.462     &      2.34  &      2.28  \\
0.474     &      1.43 &      1.37 \\
0.482     &      0.89  &      0.85   \\
0.490     &      0.43 &      0.41  \\
0.494     &      0.22  &      0.21  \\
0.498     &      0.05   &      0.04 \\

\hline
\hline
\end{tabular}
\end{tabular}
\end{center}
\end{table}
\end{document}